\documentclass[aps,a4paper,superscriptaddress,preprintnumbers,showpacs,amsmath,amssymb]{revtex4}
\usepackage{graphicx}
\usepackage{dcolumn}
\usepackage{color}
\usepackage{latexsym,amsfonts}
\usepackage{textcomp}
\usepackage{bm}

\usepackage{ulem}

\def\beq{\begin{equation}}
\def\eeq{\end{equation}}

\begin{document}

\title{Standard Electroweak Interactions\\ in Gravitational Theory
  with Chameleon Field and Torsion}
\author{A. N. Ivanov}\email{ivanov@kph.tuwien.ac.at}
\affiliation{Atominstitut, Technische Universit\"at Wien, Stadionallee
  2, A-1020 Wien, Austria}
\author{M. Wellenzohn}\email{max.wellenzohn@gmail.com}
\affiliation{Atominstitut, Technische Universit\"at Wien, Stadionallee
  2, A-1020 Wien, Austria}
\affiliation{FH Campus Wien, University of Applied Sciences, 
Favoritenstra\ss e 226, 1100 Wien, Austria}

\date{\today}

\begin{abstract}
We propose a version of a gravitational theory with the torsion field,
induced by the chameleon field. Following Hojman {\it et al.}
Phys. Rev. D {\bf 17}, 3141 (1976) the results, obtained in
Phys. Rev. D {\rm 90}, 045040 (2014), are generalised by extending the
Einstein gravity to the Einstein--Cartan gravity with the torsion
field as a gradient of the chameleon field through a modification of
local gauge invariance of minimal coupling in the Weinberg--Salam
electroweak model. The contributions of the chameleon (torsion) field
to the observables of electromagnetic and weak processes are
calculated. Since in our approach the
  chameleon--photon coupling constant $\beta_{\gamma}$ is equal to the
  chameleon--matter coupling constant $\beta$, i.e. $\beta_{\gamma} =
  \beta$, the experimental constraints on $\beta$, obtained in
  terrestrial laboratories by T. Jenke {\it et al.}
  (Phys. Rev. Lett. {\bf 112}, 115105 (2014)) and by H. Lemmel {\it et
    al.} (Phys. Lett. B {\bf 743}, 310 (2015)), can be used for the
  analysis of astrophysical sources of chameleons, proposed by
  C. Burrage {\it et al.} (Phys. Rev. D {\bf 79}, 044028 (2009)),
  A.-Ch. Davis {\it et al.} (Phys. Rev. D {\bf 80}, 064016 (2009) and
  in references therein, where chameleons induce photons because of
  direct chameleon--photon transitions in the magnetic fields.
\end{abstract}

 \maketitle

\section{Introduction}
\label{sec:introduction}

The chameleon field, the properties of which are analogous to a
quintessence \cite{Zlatev1999,Tsujikawa2013}, i.e. a canonical scalar
field invented to explain the late--time acceleration of the Universe
expansion \cite{Perlmutter1997,Riess1998,Perlmutter1999}, has been
proposed in \cite{Chameleon1,Chameleon2,Waterhouse}. In order to avoid
the problem of violation of the equivalence principle \cite{Will1993}
a chameleon mass depends on a mass density $\rho$ of a local
environment \cite{Chameleon1,Chameleon2,Waterhouse}.  The
self--interaction of the chameleon field and its interaction to a
local environment with a mass density $\rho$ are described by the
effective potential $V_{\rm eff}(\phi)$
\cite{Chameleon1,Chameleon2,Waterhouse,Brax2011,Ivanov2013,Jenke2014}
\begin{eqnarray}\label{eq:1}
	V_{\rm eff}(\phi) = V(\phi) + \rho\,e^{\,\beta
          \phi/M_{\rm Pl}},
\end{eqnarray}
where $\phi$ is a chameleon field, $\beta$ is a
chameleon--matter field coupling constant and $M_{\rm Pl} =
1/\sqrt{8\pi G_N} = 2.435\times 10^{27}\,{\rm eV}$ is the reduced
Planck mass \cite{PDG2014}. The potential $V(\phi)$ defines
self--interaction of a chameleon field.

As has been pointed out in Ref. \cite{Brax2011,Ivanov2013,Jenke2014},
ultracold neutrons (UCNs), bouncing in the gravitational field of the
Earth above a mirror and between two mirrors, can be a good laboratory
for testing of a chameleon--matter field interaction. Using the
solutions of equations of motion for a chameleon field, confined
between two mirrors, there has been found the upper limit for the
coupling constant $\beta < 5.8 \times 10^8$ \cite{Jenke2014}, which
was estimated from the contribution of a chameleon field to the
transition frequencies of the quantum gravitational states of UCNs,
bouncing in the gravitational field of the Earth. For the analysis of
the chameleon--matter field interactions in
Refs.\cite{Brax2011,Ivanov2013,Jenke2014} the potential $V(\phi)$ of a
chameleon--field self--interaction has been taken in the form of the
Ratra--Peebles potential \cite{Ratra1988} (see also
\cite{Chameleon1,Chameleon2})
\begin{eqnarray}\label{eq:2}
	V(\phi) = \Lambda^4 + \frac{\Lambda^{4 + n}}{\phi^n},
\end{eqnarray}
where $\Lambda = \sqrt[4]{3\Omega_{\Lambda} {\rm H}^2_0 M^2_{\rm Pl}}
= 2.24(2) \times 10^{-3}\,{\rm eV}$ \cite{Brax2013} with
$\Omega_{\Lambda} = 0.685^{+0.017}_{-0.016}$ and ${\rm H}_0 =
1.437(26) \times 10^{-33}\,{\rm eV}$ are the relative dark energy
density and the Hubble constant \cite{PDG2014}, respectively, and $n$
is the Ratra--Peebles index. The runaway form $\Lambda^{4 + n
}/\phi^n$ for $\phi \to \infty$ is required by the quintessence models
\cite{Zlatev1999,Tsujikawa2013}.  Such a potential of a
self--interaction of the chameleon field allows to realise the regime
of the strong chameleon--matter coupling constant $\beta \gg
10^5$ \cite{Brax2011,Ivanov2013,Jenke2014}.

Recently \cite{Ivanov2014} some new chameleon--matter field
interactions have been derived from the non--relativistic
approximation of the Dirac equation for slow fermions, moving in
spacetimes with a static metric, caused by the weak gravitational
field of the Earth and a chameleon field. The derivation of the
non--relativistic Hamilton operator of the Dirac equation has been
carried out by using the standard Foldy--Wouthuysen (SFW)
transformation. There has been also shown that the chameleon field can
serve as a source of a torsion field and torsion--matter interactions
\cite{Hehl1976}--\cite{Kostelecky2011}. A relativistic covariant
torsion--neutron interaction has been found in the following form
\cite{Ivanov2014}
\begin{eqnarray}\label{eq:3}
	{\cal L}_{\cal T}(x) = \frac{i}{2}g_{\cal T}\,{\cal
          T}_{\mu}(x)\bar{\psi}(x)\,\sigma^{\mu\nu}
        \overleftrightarrow{\partial_{\nu}}\psi(x),
\end{eqnarray}
where ${\cal T}_{\mu}$ is the torsion field, $\psi(x)$ is the neutron
field operator, $A(x)\overleftrightarrow{\partial_{\nu}}B(x) = A(x)
\partial_{\nu}B(x) - (\partial_{\nu}A(x)) B(x) $ and $\sigma^{\mu\nu}
= \frac{i}{2}(\gamma^{\mu}\gamma^{\nu} - \gamma^{\nu} \gamma^{\mu})$
is one of the Dirac matrices \cite{Itzykson1980}. In the
non--relativistic limit we get
\begin{eqnarray}\label{eq:4}
\delta {\cal L}_{\cal T}(x) = i\,g_{\cal T}\vec{\cal T}\cdot
\bar{\psi}(x)(\vec{\Sigma}\times \vec{\nabla}\,)\psi(x) + \ldots =
i\,g_{\cal T} \vec{\cal T}\cdot \varphi^{\dagger}(x)(\vec{\sigma}\times
\vec{\nabla}\,)\varphi(x) + \ldots\,,
\end{eqnarray}
where $\varphi(x)$ is the
operator of the large component of the Dirac bispinor field operator
$\psi(x)$ and $\vec{\Sigma} = \gamma^0\vec{\gamma}\gamma^5$ is the
diagonal Dirac matrix with elements $\vec{\Sigma} = {\rm
  diag}(\vec{\sigma}, \vec{\sigma}\,)$ and $\vec{\sigma}$ are the
$2\times 2$ Pauli matrices \cite{Itzykson1980}.  As has been found in
\cite{Ivanov2014} the product $g_{\cal T}\vec{\cal T}$ is equal to
\begin{eqnarray}\label{eq:5}
g_{\cal T}\vec{\cal T}(x) = - \frac{1 + 2\gamma}{4 m}\,
\frac{\beta}{M_{\rm Pl}}\,\vec{\nabla}\phi(\vec{r}\,)\,,
\end{eqnarray}
where $\gamma = 1$ for the Schwarzschild metric of a weak
gravitational field \cite{Ivanov2014}( see also
\cite{Fischbach1981,Jentschura2013,Jentschura2014}).

Following \cite{Hojman1978} we introduce the torsion field tensor
${{\cal T}^{\alpha}}_{\mu\nu}$ as follows
\begin{eqnarray}\label{eq:6}
f{{\cal T}^{\alpha}}_{\mu\nu} = {\delta^{\alpha}}_{\nu} f_{,\mu} -
{\delta^{\alpha}}_{\mu}f_{,\nu},
\end{eqnarray}
where $f_{,\mu} = \partial_{\mu}f$ and $f = e^{\,\phi_{\rm H}}$. Such
an expression one obtains from the requirement of local gauge
invariance of the electromagnetic field strength \cite{Hojman1978}
(see also section {\ref{sec:photon}}). According to \cite{Ivanov2014},
the scalar field $\phi_{\rm H}$ can be identified with a chameleon
field $\phi$ as $\phi_{\rm H} = \beta \phi/M_{\rm Pl}$. As a result,
we get
\begin{eqnarray}\label{eq:7}
{{\cal T}^{\alpha}}_{\mu\nu} = \frac{\beta}{M_{\rm
    Pl}}({\delta^{\alpha}}_{\nu}\phi_{,\mu} -
{\delta^{\alpha}}_{\mu}\phi_{,\nu}) =
g^{\alpha\lambda}\frac{\beta}{M_{\rm
    Pl}}(g_{\lambda\nu}\phi_{,\mu} -
  g_{\lambda\mu}\phi_{,\nu}) = g^{\alpha\lambda}{\cal T}_{\lambda\mu\nu},
\end{eqnarray}
where ${\delta^{\alpha}}_{\nu} = g^{\alpha\lambda}g_{\lambda\nu}$ and
$g_{\nu\lambda}$ and $g^{\lambda\alpha}$ are the metric and inverse
metric tensor, respectively. The torsion tensor field ${{\cal
    T}^{\alpha}}_{\mu\nu}$ is anti--symmetric ${T^{\alpha}}_{\mu\nu} =
- {{\cal T}^{\alpha}}_{\nu\mu}$. 

For the subsequent analysis we need a definition of the covariant
derivative $V_{\mu;\nu}$ of a vector field $V_{\mu}$ in the curve
spacetime. It is given by \cite{Feynman1995,Fliessbach2006,Rebhan2012}
\begin{eqnarray}\label{eq:8}
V_{\mu;\nu} = V_{\mu,\nu} - V_{\alpha}{\Gamma^{\alpha}}_{\mu\nu},
\end{eqnarray}
where ${\Gamma^{\alpha}}_{\mu\nu}$ is the affine connection,
determined by \cite{Hojman1978}
\begin{eqnarray}\label{eq:9}
{\Gamma^{\alpha}}_{\mu\nu} = \{{^{\alpha}}_{\mu\nu}\} - \frac{1}{2}\,
g^{\alpha\sigma}({\cal T}_{\sigma\mu\nu} - {\cal T}_{\mu\sigma\nu} -
{\cal T}_{\nu\sigma\mu}) = \{{^{\alpha}}_{\mu\nu}\} +
\frac{\beta}{M_{\rm
    Pl}}\,g^{\alpha\sigma}(- g_{\sigma\nu}\phi_{,\mu} +
g_{\mu\nu}\phi_{,\sigma}),
\end{eqnarray}
where $\{{^{\alpha}}_{\mu\nu}\}$ are the Christoffel symbols
\cite{Feynman1995,Fliessbach2006,Rebhan2012}
\begin{eqnarray}\label{eq:10}
\{{^{\alpha}}_{\mu\nu}\} =
\frac{1}{2}g^{\alpha\lambda}(g_{\lambda\mu,\nu} + g_{\lambda\nu,\mu} -
g_{\mu\nu,\lambda})
\end{eqnarray}
and ${{\cal T}^{\alpha}}_{\mu\nu}$ is the torsion tensor field
\cite{Hehl1976,Hojman1978}
\begin{eqnarray}\label{eq:11}
{{\cal T}^{\alpha}}_{\mu\nu} = {\Gamma^{\alpha}}_{\nu\mu} -
{\Gamma^{\alpha}}_{\mu\nu} = \frac{\beta}{M_{\rm
    Pl}}({\delta^{\alpha}}_{\nu}\phi_{,\mu} -
{\delta^{\alpha}}_{\mu}\phi_{,\nu}).
\end{eqnarray}
We introduce the contribution of the torsion field in agreement with
Hojman {\it et al.} \cite{Hojman1978} (see Eq.(38) of
Ref.\cite{Hojman1978}). 

Having determined the affine connection we may introduce the
Riemann--Christoffel tensor ${{\cal R}^{\alpha}}_{\mu\nu\lambda}$ or
curvature tensor as \cite{Feynman1995,Fliessbach2006,Rebhan2012}
\begin{eqnarray}\label{eq:12}
{{\cal R}^{\alpha}}_{\mu\nu\lambda} = {\Gamma^{\alpha}}_{\mu\nu,\lambda} -
{\Gamma^{\alpha}}_{\mu\lambda,\nu} +
{\Gamma^{\alpha}}_{\lambda\varphi}{\Gamma^{\varphi}}_{\mu\nu} -
{\Gamma^{\alpha}}_{\nu\varphi}{\Gamma^{\varphi}}_{\mu\lambda},
\end{eqnarray}
which is necessary for the definition of the Lagrangian of the
gravitational field in terms of the scalar curvature ${\cal R}$
\cite{Feynman1995} related to the
Riemann--Christoffel tensor ${{\cal R}^{\alpha}}_{\mu\nu\lambda}$ by
\cite{Feynman1995,Fliessbach2006,Rebhan2012}
\begin{eqnarray}\label{eq:13}
{\cal R} = g^{\mu\lambda}{{\cal R}^{\alpha}}_{\mu\alpha\lambda} =
g^{\mu\lambda}{\cal R}_{\mu \lambda}.
\end{eqnarray}
Here ${\cal R}_{\mu \lambda}$ is the Ricci tensor
\cite{Feynman1995,Fliessbach2006,Rebhan2012}.  Following
\cite{Hojman1978} and skipping intermediate calculations one may show
that the scalar curvature ${\cal R}$ is equal to
\begin{eqnarray}\label{eq:14}
{\cal R} = R + \frac{3\beta^2}{M^2_{\rm
    Pl}}\,g^{\mu\nu}\phi_{,\mu}\phi_{,\nu},
\end{eqnarray}
where the curvature $R$ is determined by the Riemann--Christoffel
tensor Eq.(\ref{eq:12}) with the replacement
${\Gamma^{\alpha}}_{\mu\nu} \to \{{^{\alpha}}_{\mu\nu}\}$.

The paper is organised as follows. In section \ref{sec:chameleon} we
consider the chameleon field in the gravitational field with torsion
(a version of the Einstein--Cartan gravity), caused by the chameleon
field. We derive the effective Lagrangian and the equations of motion
of the chameleon field coupled to the gravitational field (a version
of the Einstein gravity with a scalar self--interacting field). In
section \ref{sec:photon} we analyse the interaction of the chameleon
(torsion) field with the electromagnetic field, coupled also to the
gravitational field. Following Hojman {\it et al.}  \cite{Hojman1978}
and modifying local gauge invariance of the electromagnetic strength
tensor field we derive the torsion field tensor ${{\cal
    T}^{\alpha}}_{\mu\nu}$ in terms of the chameleon field (see
Eq.(\ref{eq:7}). In section \ref{sec:gamma} we analyse the torsion
(chameleon) - photon interactions in terms of the two--photon decay
$\phi \to \gamma + \gamma$ of the chameleon and the photon--chameleon
scattering $\gamma + \phi \to \phi + \gamma$. We show that the
amplitudes of the two--photon decay and the photon--chameleon
scattering are gauge invariant. In order words we show that the
replacement of the photon polarisation vectors by their 4--momenta
leads to the vanishing of the amplitudes of the two--photon decay and
the photon--chameleon scattering. In section \ref{sec:higgs} we
investigate the Weinberg--Salam electroweak model \cite{PDG2014}
without fermions. We derive the effective Lagrangian of the
electroweak bosons, the electromagnetic field and the Higgs boson
coupled to the gravitational and chameleon field. Such a derivation we
carry out by means of a modification of local gauge invariance. In
section \ref{sec:fermion} we include fermions into the Weinberg--Salam
model and derive the effective interactions of the electroweak bosons,
the Higgs field and fermions with the gravitational and chameleon
field. In section \ref{sec:beta} we calculate the contributions of the
chameleon to the charge radii of the neutron and proton. We calculate
the contributions of the chameleon to the correlation coefficients of
the neutron $\beta^-$--decay $n \to p + e^- + \bar{\nu}_e + \phi$ with
a polarised neutron and unpolarised proton and electron. In addition
we calculate the cross section for the neutron $\beta^-$--decay $\phi
+ n \to p + e^- + \bar{\nu}_e$, induced by the chameleon field. In
section \ref{sec:conclusion} we discuss the obtained results and
perspectives of the experimental analysis of the approach, developed
in this paper, and of observation of the neutron $\beta^-$--decay,
induced by the chameleon.

\section{Torsion gravity and effective Lagrangian of 
chameleon field}
\label{sec:chameleon}

The action of the gravitational field with torsion, the chameleon
field and matter fields we define by \cite{Chameleon1,Chameleon2}
\begin{eqnarray}\label{eq:15}
S_{\rm g,ch} = \int d^4x\,\sqrt{-g}\,{\cal L}[{\cal R},\phi] + \int
d^4x\,\sqrt{-\tilde{g}}\,{\cal L}_m[\tilde{g}_{\mu\nu}],
\end{eqnarray}
where the Lagrangian ${\cal L}[{\cal R}, \phi]$ is given by
\begin{eqnarray}\label{eq:16}
{\cal L}[{\cal R},\phi] = \frac{1}{2}\,M^2_{\rm Pl}\,{\cal R} +
\frac{1}{2}\,(1 - 3\,\beta^2)\,\phi_{,\mu}\phi^{,\mu} - V(\phi).
\end{eqnarray}
Here $\phi_{,\mu} = \partial \phi/\partial x^{\mu}$ and $\phi^{,\mu} =
\partial \phi/\partial x_{\mu}$ and $V(\phi)$ is the potential of the
self--interaction of the chameleon field Eq.(\ref{eq:2}). The matter
fields are described by the Lagrangian ${\cal
  L}_m[\tilde{g}_{\mu\nu}]$. The interaction of the matter field with
the chameleon field runs through the metric tensor
$\tilde{g}_{\mu\nu}$ in the Jordan--frame
\cite{Chameleon1,Chameleon2,Fujii2004,Capozziello2010}, which is
conformally related to the Einstein--frame metric tensor $g_{\mu\nu}$
by $\tilde{g}_{\mu\nu}= f^2\,g_{\mu\nu}$ (or $\tilde{g}^{\mu\nu} =
f^{-2}\,g^{\mu\nu}$) and $\sqrt{- \tilde{g}} = f^4 \,\sqrt{-g}$ with
$f = e^{\,\beta\phi/M_{\rm Pl}}$ \cite{Chameleon1,Dicke1962}. The
factor $e^{\,\beta\phi/M_{\rm Pl}}$ can be interpreted also as a
conformal coupling to matter fields \cite{Chameleon1,Chameleon2}.
Using Eq.(\ref{eq:14}) we transcribe the action Eq.(\ref{eq:15}) into
the form
\begin{eqnarray}\label{eq:17}
S_{\rm g,ch} = \int d^4x\,\sqrt{-g}\,\Big(\frac{1}{2}\,M^2_{\rm Pl}\,R
+ {\cal L}[\phi]\Big) +
\int d^4x\,\sqrt{-\tilde{g}}\,{\cal L}_m[\tilde{g}_{\mu\nu}],
\end{eqnarray}
where the contribution of the torsion field to the scalar curvature is
absorbed by the kinetic term of the chameleon field. The Lagrangian
${\cal L}[\phi]$ is equal to
\begin{eqnarray}\label{eq:18}
{\cal L}[\phi] = \frac{1}{2}\,\phi_{,\mu}\phi^{,\mu} - V(\phi).
\end{eqnarray}
The total Lagrangian in the action Eq.(\ref{eq:17}) is usually
referred as the Lagrangian in the Einstein frame, where $g_{\mu\nu}$
as well as $g^{\mu\nu}$ is the Einstein--frame metric such as
$\phi_{,\mu}\phi^{,\mu} = g^{\mu\nu}\phi_{,\mu}\phi_{,\nu}$ and
$\Box\phi = (1/\sqrt{-g})(\sqrt{-g}\,\phi^{,\mu})_{;\mu} =
(1/\sqrt{-g})(\sqrt{-g}\,g^{\mu\nu}\phi_{,\nu})_{;\mu}$
\cite{Dicke1962} (see also \cite{Chameleon1,Chameleon2}).

Varying the action Eq.(\ref{eq:17}) with respect to $\phi_{,\mu}$ and
$\phi$ we arrive at the equation of motion of the chameleon field
\begin{eqnarray}\label{eq:19}
\frac{\partial}{\partial x^{\mu}}\frac{\delta (\sqrt{- g}\,{\cal
    L}[\phi])}{\delta \phi_{,\mu}} = \frac{\delta (\sqrt{- g}\,{\cal
    L}[\phi])}{\delta \phi} + \frac{\delta (\sqrt{- \tilde{g}}\,{\cal
    L}_m)}{\delta \phi},
\end{eqnarray}
Using the Lagrangian Eq.(\ref{eq:18}) we transform Eq.(\ref{eq:19})
into the form
\begin{eqnarray}\label{eq:20}
\frac{1}{\sqrt{ - g}}\,\frac{\partial}{\partial
  x^{\mu}}(\sqrt{-g}\,\phi^{,\mu}) = - V'_{\phi}(\phi)-
f'_{\phi}\,f^3\, \tilde{g}^{\alpha\lambda}\,\frac{2}{\sqrt{ -
    \tilde{g}}}\frac{\delta (\sqrt{- \tilde{g}}\,{\cal L}_m)}{\delta
  \tilde{g}^{\alpha\lambda}},
\end{eqnarray}
where $V'_{\phi}(\phi)$ and $f'_{\phi}$ are derivatives with respect
to $\phi$.  Since by definition \cite{Chameleon1,Chameleon2} the
derivative
\begin{eqnarray}\label{eq:21}
\frac{2}{\sqrt{ - \tilde{g}}}\frac{\delta (\sqrt{- \tilde{g}}\,{\cal
    L}_m)}{\delta \tilde{g}^{\alpha\lambda}} =
\tilde{T}_{\alpha\lambda}
\end{eqnarray}
is a matter stress--energy tensor in the Jordan frame,
Eq.(\ref{eq:20}) takes the form
\begin{eqnarray}\label{eq:22}
\Box \phi = - V'_{\phi}(\phi) -
f'_{\phi}\,f^3\,{\tilde{T}^{\alpha}}_{\alpha},
\end{eqnarray}
where ${\tilde{T}^{\alpha}}_{\alpha} =
\tilde{g}^{\alpha\lambda}\,\tilde{T}_{\alpha\lambda}$. For a
pressureless matter ${\tilde{T}^{\alpha}}_{\alpha} = \tilde{\rho}$,
where $\tilde{\rho}$ is a matter density in the Jordan frame, related
to a matter density in the Einstein frame $\rho$ by $\tilde{\rho}
=f^{-3}\,\rho$ \cite{Chameleon1} we get
\begin{eqnarray}\label{eq:23}
\Box \phi = - V'_{\phi}(\phi) - \rho\,f'_{\phi},
\end{eqnarray}
where we have set $f^3\,{\tilde{T}^{\alpha}}_{\alpha} = \rho$
\cite{Chameleon1}. Then, $V'_{\phi}(\phi) - \rho\,f'_{\phi}$ coincides
with the derivative of the effective potential of the
chameleon--matter interaction $V_{\rm eff}(\phi)$ with respect to
$\phi$, given by Eq.(\ref{eq:1}) for $f = e^{\,\beta\phi/M_{\rm Pl}}$.

\section{Torsion gravity with chameleon and electromagnetic fields}
\label{sec:photon}

In this section we analyse the interactions of the torsion (chameleon)
field with the electromagnetic field. The action of the gravitational
field, the chameleon field, the matter fields and the electromagnetic
field is equal to
\begin{eqnarray}\label{eq:24}
S_{\rm g,ch,em} &=& \int d^4x\,\sqrt{-g}\,\Big(\frac{1}{2}\,M^2_{\rm
  Pl}\,R + \frac{1}{2}\,\phi_{,\mu}\phi^{,\mu} - V(\phi)\Big) -
\frac{1}{4} \int d^4x\,\sqrt{-\tilde{g}}\,\tilde{g}^{\alpha\mu}
\tilde{g}^{\beta\nu}{\cal F}_{\alpha\beta} {\cal
  F}_{\mu\nu}\nonumber\\ &+& \int d^4x\,\sqrt{-\tilde{g}}\,{\cal
  L}_m[\tilde{g}_{\mu\nu}],
\end{eqnarray}
where $\tilde{g}_{\mu\nu} = f^2\,g_{\mu\nu}$. Since $\sqrt{-
  \tilde{g}} = f^4 \,\sqrt{-g}$, we get that
$\sqrt{-\tilde{g}}\,\tilde{g}^{\alpha\mu} \tilde{g}^{\beta\nu} =
\sqrt{-g}\,g^{\alpha\mu}g^{\beta\nu}$. The term
$\sqrt{-\tilde{g}}\,{\cal L}_m[\tilde{g}_{\mu\nu}]$ describes an
environment where the chameleon field couples to the electromagnetic
field.

Following then Hojman {\it et al.}  \cite{Hojman1978} we define the
electromagnetic strength tensor field ${\cal F}_{\mu\nu}$ in the
gravitational and torsion field
\begin{eqnarray}\label{eq:25}
{\cal F}_{\mu\nu} = A_{\nu;\mu} - A_{\mu;\nu} = A_{\nu,\mu} -
A_{\mu,\nu} - A_{\alpha}{{\cal T}^{\alpha}}_{\mu\nu} = F_{\mu\nu} -
A_{\lambda}{{\cal T}^{\lambda}}_{\mu\nu},
\end{eqnarray}
where $F_{\mu\nu} = A_{\nu,\mu} - A_{\mu,\nu}$ and $A_{\mu}$ is the
electromagnetic 4--potential.  According to Hojman {\it et al.}
\cite{Hojman1978}, under a gauge transformation the electromagnetic
potential transforms as follows
\begin{eqnarray}\label{eq:26}
A_{\mu} \to A'_{\mu} = A_{\mu} + {c_{\mu}}^{\alpha}(\Phi)\Lambda_{,\alpha},
\end{eqnarray}
where ${c_{\mu}}^{\alpha}(\Phi)$ is a functional of the scalar field
$\Phi$, which we identify with the chameleon field $\Phi =
\beta\phi/M_{\rm Pl}$, i.e.  ${c_{\mu}}^{\alpha}(\Phi) \to
           {c_{\mu}}^{\alpha}(\phi) =
           {\delta_{\mu}}^{\alpha}e^{\,\beta\phi/M_{\rm Pl}}$, and
           $\Lambda$ is an arbitrary gauge function. The gauge
           invariance of the electromagnetic field strength imposes
           the constraint \cite{Hojman1978}
\begin{eqnarray}\label{eq:27}
{c_{\nu}}^{\alpha}(\phi)_{,\mu} - {c_{\mu}}^{\alpha}(\phi)_{,\nu} -
{c_{\sigma}}^{\alpha}(\phi){{\cal T}^{\sigma}}_{\mu\nu} = 0.
\end{eqnarray}
This gives the torsion tensor field given by Eq.(\ref{eq:6}) and
Eq.(\ref{eq:7}).  Substituting Eq.(\ref{eq:25}) into Eq.(\ref{eq:24})
we arrive at the expression
\begin{eqnarray}\label{eq:28}
S_{\rm g,ch, em} &=& \int d^4x\,\sqrt{-g}\,\Big( 
\frac{1}{2}\,M^2_{\rm Pl}\,R + \frac{1}{2}\,\phi_{,\mu}\phi^{,\mu} -
V(\phi) - \frac{1}{4}\,F_{\mu\nu} F^{\mu\nu}\nonumber\\ &+&
\frac{1}{2}\,g^{\alpha\mu} g^{\beta\nu} F_{\mu\nu} A_{\sigma}{{\cal
    T}^{\sigma}}_{\alpha\beta} - \frac{1}{4}\,g^{\alpha\mu}
g^{\beta\nu}A_{\sigma}A_{\lambda}{{\cal
    T}^{\sigma}}_{\alpha\beta}{{\cal T}^{\lambda}}_{\mu\nu}\Big) +
\int d^4x\,\sqrt{-\tilde{g}}\,{\cal L}_m[\tilde{g}_{\mu\nu}].
\end{eqnarray}
Using the definition of the torsion tensor field Eq.(\ref{eq:7}) we
arrive at the action
\begin{eqnarray}\label{eq:29}
S_{\rm g,ch, em} &=& \int d^4x\,\sqrt{-g}\,\Big( 
\frac{1}{2}\,M^2_{\rm Pl}R + \frac{1}{2}\,\phi_{,\mu}\phi^{,\mu} -
V(\phi) - \frac{1}{4}\,F^{\mu\nu} F_{\mu\nu} -
\frac{1}{2}\,\frac{\beta}{M_{\rm Pl}}\, F^{\mu\nu} (A_{\mu}\phi_{,\nu}
- A_{\nu}\phi_{,\mu})\nonumber\\ &-&
\frac{1}{4}\,\frac{\beta^2}{M^2_{\rm Pl}}\,(A^{\mu}\phi^{,\nu} -
A^{\nu}\phi^{,\mu})(A_{\mu}\phi_{,\nu} - A_{\nu}\phi_{,\mu})\Big) +
\int d^4x\,\sqrt{-\tilde{g}}\,{\cal L}_m[\tilde{g}_{\mu\nu}]
\end{eqnarray}
Thus, because of the torsion--electromagnetic field interaction the
chameleon becomes unstable under the two--photon decay $\phi \to
\gamma + \gamma$ and may scatter by photons $ \gamma + \phi \to \phi +
\gamma$ with the chameleon--matter coupling constant $\beta/M_{\rm
  Pl}$.  These reactions are described by the effective Lagrangians
\begin{eqnarray}\label{eq:30}
{\cal L}_{\phi \gamma\gamma} = - \sqrt{-g}\,\frac{\beta}{M_{\rm
    Pl}}\,F^{\mu\nu} A_{\mu}\phi_{,\nu} 
\end{eqnarray}
and
\begin{eqnarray}\label{eq:31}
{\cal L}_{\phi \phi \gamma\gamma} &=&-\,\sqrt{-g}\,
\frac{1}{4}\,\frac{\beta^2}{M^2_{\rm Pl}}\,(A^{\mu}\phi^{,\nu} -
A^{\nu}\phi^{,\mu})(A_{\mu}\phi_{,\nu} - A_{\nu}\phi_{,\mu}) = \nonumber\\ &=&
-\,\sqrt{-g}\,\frac{1}{2}\,\frac{\beta^2}{M^2_{\rm
    Pl}}\,(A_{\mu}A^{\mu}\phi_{,\nu}\phi^{,\nu} -
A_{\mu}A^{\nu}\phi^{,\mu} \phi_{,\nu}).
\end{eqnarray}
For the application of the action Eq.(\ref{eq:29}) with
chameleon--photon interactions to the calculation of the specific
reactions of the chameleon--photon and chameleon--photon--matter
interactions we have to fix the gauge of the electromagnetic
field. We may do this in a standard way \cite{Itzykson1980}
\begin{eqnarray}\label{eq:32}
S_{\rm g,ch, em} &=& \int d^4x\,\sqrt{-g}\,\Big( 
\frac{1}{2}\,M^2_{\rm Pl}\,R + \frac{1}{2}\,\phi_{,\mu}\phi^{,\mu} -
V(\phi) - \frac{1}{4}\,{\cal F}_{\mu\nu} {\cal F}^{\mu\nu} +
f(A^{\mu}_{;\mu}, A_{\nu}, \phi)\Big)\nonumber\\
 &+& \int
d^4x\,\sqrt{-\tilde{g}}\,{\cal L}_m[\tilde{g}_{\mu\nu}],
\end{eqnarray}
where $f(A^{\mu}_{;\mu}, A_{\nu}, \phi)$ is a gauge fixing functional
and the divergence ${A^{\mu}}_{;\mu}$ is defined by
\cite{Fliessbach2006,Rebhan2012}
\begin{eqnarray}\label{eq:33}
{A^{\mu}}_{;\mu} = {A^{\mu}}_{,\mu} + {\tilde{\Gamma}^{\mu}}_{\mu\nu} A^{\nu}.
\end{eqnarray}
The affine connection ${\tilde{\Gamma}^{\alpha}}_{\mu\nu}$ we have to
calculate for the Jordan--frame metric $\tilde{g}_{\mu\nu} =
f^2\,g_{\mu\nu}$ \cite{Chameleon1,Fujii2004}. We get
\begin{eqnarray}\label{eq:34}
{\tilde{\Gamma}^{\alpha}}_{\mu\nu} = \{{^{\alpha}}_{\mu\nu}\} +
{\delta^{\alpha}}_{\mu}\,f^{-1}\,f_{,\nu} = \{{^{\alpha}}_{\mu\nu}\} +
\frac{\beta}{M_{\rm Pl}}\,{\delta^{\alpha}}_{\mu}\,\phi_{,\nu}
\end{eqnarray}
such as ${\tilde{\Gamma}^{\alpha}}_{\nu\mu} -
{\tilde{\Gamma}^{\alpha}}_{\mu\nu} = {{\cal T}^{\alpha}}_{\mu\nu}$
(see Eq.(\ref{eq:7})).  As a result, the divergence $A^{\mu}_{;\mu}$
is equal to \cite{Fliessbach2006,Rebhan2012}
\begin{eqnarray}\label{eq:35}
A^{\mu}_{;\mu} = \frac{1}{\sqrt{- g}}\frac{\partial(\sqrt{- g}\,A^{\mu})}
{\partial x^{\mu}} + 4\,\frac{\beta}{M_{\rm Pl}}\,\phi_{,\nu}A^{\nu}.
\end{eqnarray}
Since a gauge condition should not depend on the chameleon field, we
propose to fix a gauge as follows
\begin{eqnarray}\label{eq:36}
f(A^{\mu}_{;\mu}, A_{\nu}, \phi) = -
\frac{1}{2\xi}\,\Big(A^{\mu}_{;\mu} - 4\, \frac{\beta}{M_{\rm
    Pl}}\,\phi_{,\nu} A^{\nu}\Big)^2,
\end{eqnarray}
where $\xi$ is a gauge parameter.  Now we are able to investigate some
specific processes of chameleon--photon interactions.

\section{Chameleon--photon interactions}
\label{sec:gamma}

The specific processes of the chameleon--photon interaction, which we
analyse in this section, are i) the two--photon decay $\phi \to \gamma
+ \gamma$ and ii) the photon--chameleon scattering $\gamma + \phi \to
\phi + \gamma$. The calculation of these reactions we carry out in the
Minkowski spacetime. For this aim in the interactions Eq.(\ref{eq:30})
and Eq.(\ref{eq:31}) we make a replacement $g^{\mu\nu} \to
\eta^{\mu\nu}$, where $\eta^{\mu\nu}$ is the metric tensor in the
Minkowski space time with only diagonal components $(+1, -1,-1,-1)$,
and $g \to {\rm det}\{\eta_{\mu\nu}\} = -1$.

\subsection{Two--photon $\phi \to
 \gamma + \gamma$ decay of the chameleon}

\begin{figure}
\centering \includegraphics[width=0.20\linewidth]{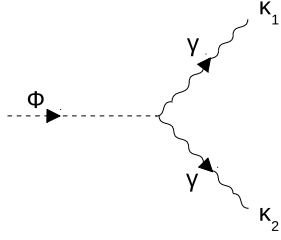}
\caption{Feynman diagram for the $\phi \to \gamma + \gamma$ decay.}
 \label{fig:torsion}
\end{figure}

For the calculation of the two--photon decay rate of the chameleon we
use the Lagrangian Eq.(\ref{eq:30}). The Feynman diagram of the
amplitude of the two--photon decay of the chameleon is shown in
Fig.\,{\ref{fig:torsion}}. The analytical expression of the amplitude
of the $\phi \to \gamma + \gamma$ decay is equal to
\begin{eqnarray}\label{eq:37}
M(\phi \to \gamma\,\gamma) = -\,2\,\frac{\beta}{M_{\rm
    Pl}}\,((\varepsilon^*_1\cdot \varepsilon^*_2)(k_1\cdot k_2) -
(\varepsilon^*_1\cdot k_2)(\varepsilon^*_2\cdot k_1)),
\end{eqnarray}
where $\varepsilon^*_j(k_j)$ and $k_j$ for $j = 1,2$ are the
polarisation vectors and the momenta of the decay photons, obeying the
constraints $\varepsilon^*_j(k_j)\cdot k_j = 0$. Skipping standard
calculations we obtain the following expression for the two--photon
decay rate of the chameleon
\begin{eqnarray}\label{eq:38}
\Gamma(\phi \to \gamma\,\gamma) = \frac{\beta^2}{M^2_{\rm
    Pl}}\,\frac{m^3_{\phi}}{8\pi},
\end{eqnarray}
where $m_{\phi}$ is the chameleon mass, defined by \cite{Ivanov2013}
\begin{eqnarray}\label{eq:39}
m_{\phi} = \Lambda
\sqrt{n(n+1)}\Big(\frac{\beta \rho}{n M_{\rm Pl} \Lambda^3}\Big)^{\textstyle
  \frac{n + 2}{2n + 2}}
\end{eqnarray}
as a function of the chameleon--matter coupling constant $\beta$, the
environment density $\rho$ and the Ratra--Peebles index $n$.

\begin{figure}
\centering
\includegraphics[width=0.75\linewidth]{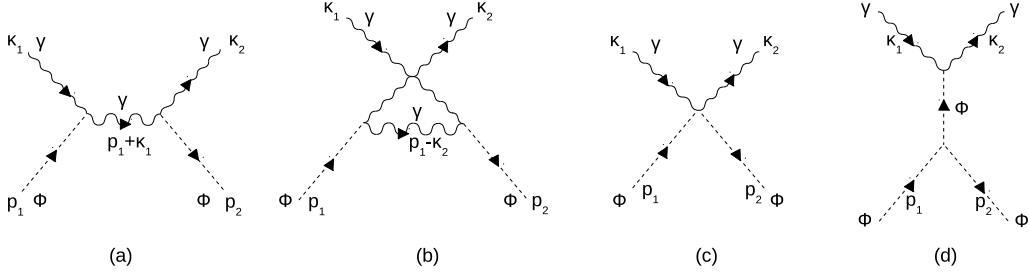}
\caption{Feynman diagram for the amplitude of the photon--chameleon
  (Compton) scattering}
 \label{fig:torsion2}
\end{figure}

\vspace{-0.3in}

\subsection{Photon--chameleon  $\gamma + \phi \to
 \phi + \gamma$ scattering }

The Feynman diagrams of the amplitude of the photon--chameleon
scattering are shown in Fig.\,{\ref{fig:torsion2}}.  The contributions
of the diagrams in Fig.\,{\ref{fig:torsion2}} are given by
\begin{eqnarray}\label{eq:40}
M^{(a)}(\gamma\,\phi\, \to \phi\,\gamma) &=& \frac{\beta^2}{M^2_{\rm
    Pl}}\,\Big(k^{\mu}_1(\varepsilon_1\cdot p_1) -
\varepsilon^{\mu}_1(k_1\cdot p_1)\Big)\,D_{\mu\alpha}(q)
\Big(k^{\alpha}_2 (\varepsilon^*_2 \cdot p_2) -
\varepsilon^{*\alpha}_2(k_2 \cdot p_2)\Big)\Big|_{q = p_1 + k_1 = p_2
  + k_2}\nonumber\\ &+&\frac{\beta^2}{M^2_{\rm
    Pl}}\,\varepsilon_{1\mu}p_{1\nu}\Big(q^{\mu}q^{\alpha}D^{\nu\beta}(q)
- q^{\mu}q^{\beta}D^{\nu\alpha}(q) - q^{\nu}q^{\alpha} D^{\mu\beta}(q)
+ q^{\nu}q^{\beta} D^{\mu\alpha}(q)\Big)
\varepsilon^*_{2\alpha}p_{2\beta}\Big|_{q = p_1 + k_1 = p_2 +
  k_2}\nonumber\\ 
&+&\frac{\beta^2}{M^2_{\rm
    Pl}}\,\Big(k^{\mu}_1(\varepsilon_1\cdot p_1) -
\varepsilon^{\mu}_1(k_1\cdot p_1)\Big)\Big(q^{\alpha}
           {D_{\mu}}^{\beta}(q) - q^{\beta}
           {D_{\mu}}^{\alpha}(q)\Big)\,\varepsilon^*_{2\alpha}p_{2\beta}\Big|_{q
             = p_1 + k_1 = p_2 + k_2
           }\nonumber\\ 
&+&\frac{\beta^2}{M^2_{\rm
               Pl}}\,\varepsilon_{1\mu} p_{1\nu}\Big(q^{\mu}{D^{\nu}}_{\alpha}(q)
             - q^{\nu}{D^{\mu}}_{\alpha}\Big)\Big(k^{\alpha}_2
             (\varepsilon^*_2\cdot p_2) -
             \varepsilon^{*\alpha}_2(k_2\cdot p_2)\Big)\Big|_{q = p_1
               + k_1 = p_2 + k_2},
\end{eqnarray}
\begin{eqnarray}\label{eq:41}
M^{(b)}(\gamma\,\phi \to \phi\,\gamma) &=& \frac{\beta^2}{M^2_{\rm
    Pl}}\,\Big(k^{\mu}_2(\varepsilon^*_2\cdot p_1) -
\varepsilon^{*\mu}_2(k_2\cdot p_1)\Big)\,D_{\mu\alpha}(q)
\Big(k^{\alpha}_1 (\varepsilon_1 \cdot p_2) -
\varepsilon^{\alpha}_1(k_1\cdot p_2)\Big)\Big|_{q = p_1 - k_2 = p_2 -
  k_1}\nonumber\\ &+&\frac{\beta^2}{M^2_{\rm
    Pl}}\,\varepsilon^*_{2\mu}p_{1\nu}\Big(q^{\mu}q^{\alpha}D^{\nu\beta}(q)
- q^{\mu}q^{\beta}D^{\nu\alpha}(q) - q^{\nu}q^{\alpha} D^{\mu\beta}(q)
+ q^{\nu}q^{\beta}
D^{\mu\alpha}(q)\Big)\,\varepsilon_{1\alpha}p_{2\beta}\Big|_{q = p_1 -
  k_2 = p_2 - k_1}\nonumber\\ &-&\frac{\beta^2}{M^2_{\rm
    Pl}}\,\Big(k^{\mu}_2(\varepsilon^*_2\cdot p_1) -
\varepsilon^{*\mu}_2(k_2\cdot p_1)\Big)\Big(q^{\alpha}
           {D_{\mu}}^{\beta}(q) - q^{\beta}
           {D_{\mu}}^{\alpha}(q)\Big)\,\varepsilon_{1\alpha}p_{2\beta}\Big|_{q
             = p_1 - k_2 = p_2 - k_1}
           \nonumber\\ &-&\frac{\beta^2}{M^2_{\rm
               Pl}}\,\varepsilon^*_{2\mu}
           p_{1\nu}\Big(q^{\mu}{D^{\nu}}_{\alpha}(q) -
           q^{\nu}{D^{\mu}}_{\alpha}\Big)\Big(k^{\alpha}_1
           (\varepsilon_1\cdot p_2) - \varepsilon^{\alpha}_1(k_1\cdot
           p_2)\Big)\Big|_{q = p_1 - k_2 = p_2 - k_1},
\end{eqnarray}
\begin{eqnarray}\label{eq:42}
M^{(c)}(\gamma\,\phi\to
             \phi\,\gamma) = \frac{\beta^2}{M^2_{\rm
                 Pl}}\,\Big(- 2(\varepsilon^*_2\cdot
             \varepsilon_1)(p_1\cdot p_2) + (\varepsilon^*_2\cdot
             p_1)(\varepsilon_1\cdot p_2) + (\varepsilon^*_2\cdot
             p_2)(\varepsilon_1\cdot
             p_1)\Big),
\end{eqnarray}
\begin{eqnarray}\label{eq:43}
 M^{(d)}(\gamma\,\phi \to \phi\,\gamma) &=& 2\,n(n + 1) (n +
 2)\,\frac{\beta}{M_{\rm Pl}}\,\frac{\Lambda^{n + 4}}{\phi^{n +
     3}_{\rm min}}\,\frac{(\varepsilon^*_2\cdot
   k_1)(\varepsilon_1\cdot k_2) - (\varepsilon^*_2\cdot
   \varepsilon_1)(k_1\cdot k_2)}{m^2_{\phi} - q^2 - i0}\Big|_{q = k_2
   - k_1 = p_1 - p_2},
\end{eqnarray}
where $\varepsilon_1$ and $\varepsilon^*_2$ are the photon
polarisation vectors in the initial and final states of the
photon--chameleon scattering. They depend on the photon momenta
$\varepsilon_1(k_1)$ and $\varepsilon^*_2(k_2)$ and obey the
constraints $\varepsilon_1(k_1)\cdot k_1 = \varepsilon^*_2(k_2)\cdot
k_2 = 0$. The chameleon field mass $m_{\phi}$ is defined by
Eq.(\ref{eq:39}). The vertex of $\phi^3$ interaction is defined by the
effective Lagrangian
\begin{eqnarray}\label{eq:44}
{\cal L}_{\phi\phi\phi} = \frac{n(n+1)(n+2)}{6}\,\frac{\Lambda^{n +
    4}}{\phi^{n + 3}_{\rm min}}\,\phi^3.
\end{eqnarray}
Here $\phi_{\rm min}$ is the minimum of the chameleon field, given by
\cite{Brax2011,Ivanov2013}
\begin{eqnarray}\label{eq:45}
\phi_{\rm min} =\Lambda\Big(\frac{n M_{\rm
    Pl}\Lambda^3}{\beta \rho}\Big)^{\textstyle \frac{1}{n + 1}},
\end{eqnarray}
where $\rho$ is the density of the medium in which the chameleon field
propagates.  The photon propagator $D_{\alpha\beta}(q)$ is equal to
\begin{eqnarray}\label{eq:46}
D_{\alpha\beta}(q) = \frac{1}{q^2 + i 0}\Big(g_{\alpha\beta} - (1 -
\xi)\frac{q_{\alpha}q_{\beta}}{q^2}\Big).
\end{eqnarray}
One may show that the amplitudes $M^{(a)}(\gamma\,\phi \to
\phi\,\gamma)$ and $M^{(b)}(\gamma\,\phi \to \phi\,\gamma)$ do not
depend on the longitudinal part of the photon propagator. As a result
the amplitudes $M^{(a)}(\gamma\,\phi \to \phi\,\gamma)$ and
$M^{(b)}(\gamma\,\phi \to \phi\,\gamma)$ can be transcribed into the
form
\begin{eqnarray}\label{eq:47}
M^{(a)}(\gamma\,\phi \to \phi\,\gamma) &=& \frac{\beta^2}{M^2_{\rm
    Pl}}\,\frac{1}{q^2 + i 0}\,\Big((\varepsilon^*_2\cdot
p_2)(\varepsilon_1\cdot p_1)(k_1\cdot k_2) - (\varepsilon^*_2\cdot
p_2)(\varepsilon_1\cdot k_2)(k_1\cdot p_1)\nonumber\\ && -
(\varepsilon^*_2\cdot k_1)(\varepsilon_1\cdot p_1)(k_2\cdot p_2) +
(\varepsilon^*_2\cdot \varepsilon_1)(k_1\cdot p_1)(k_2\cdot
p_2)\Big)\Big|_{q = p_1 + k_1 = p_2 +
  k_2}\nonumber\\ &+&\frac{\beta^2}{M^2_{\rm Pl}}\,\frac{1}{q^2 + i
  0}\,\Big((\varepsilon^*_2\cdot q)(\varepsilon_1\cdot q)(p_1\cdot
p_2) - (\varepsilon^*_2\cdot p_1)(\varepsilon_1\cdot q)(p_2\cdot
q)\nonumber\\ && - (\varepsilon^*_2\cdot q)(\varepsilon_1\cdot
p_2)(p_1\cdot q) + (\varepsilon^*_2\cdot \varepsilon_1)(p_1\cdot
q)(p_2\cdot q)\Big)\Big|_{q = p_1 + k_1 = p_2 +
  k_2}\nonumber\\ &+&\frac{\beta^2}{M^2_{\rm Pl}}\,\frac{1}{q^2 + i
  0}\,\Big((\varepsilon^*_2\cdot q)(\varepsilon_1\cdot p_1)(k_1\cdot
p_2) - (\varepsilon^*_2\cdot q)(\varepsilon_1\cdot p_2)(k_1\cdot
p_1)\nonumber\\ && - (\varepsilon^*_2\cdot k_1)(\varepsilon_1\cdot
p_1)(p_2\cdot q) + (\varepsilon^*_2\cdot \varepsilon_1)(k_1\cdot
p_1)(p_2\cdot q)\Big)\Big|_{q = p_1 + k_1 = p_2 + k_2}
\nonumber\\ &+&\frac{\beta^2}{M^2_{\rm Pl}}\,\frac{1}{q^2 + i
  0}\,\Big((\varepsilon^*_2\cdot p_2)(\varepsilon_1\cdot q)(k_2\cdot
p_1) - (\varepsilon^*_2\cdot p_2)(\varepsilon_1\cdot k_2)(p_1\cdot
q)\nonumber\\ && - (\varepsilon^*_2\cdot p_1)(\varepsilon_1\cdot
q)(k_2\cdot p_2) + (\varepsilon^*_2\cdot \varepsilon_1)(k_2\cdot
p_2)(p_1\cdot q)\Big)\Big|_{q = p_1 + k_1 = p_2 + k_2}
\end{eqnarray}
and
\begin{eqnarray}\label{eq:48}
M^{(b)}(\gamma\,\phi \to \phi\,\gamma) &=& \frac{\beta^2}{M^2_{\rm
    Pl}}\,\frac{1}{q^2 + i0}\Big((\varepsilon^*_2\cdot
p_1)(\varepsilon_1\cdot p_2)(k_1\cdot k_2) - (\varepsilon^*_2\cdot
k_1)(\varepsilon_1\cdot p_2)(k_2\cdot p_1)\nonumber\\ && -
(\varepsilon^*_2\cdot p_1)(\varepsilon_1\cdot k_2)(k_1\cdot p_2) +
(\varepsilon^*_2\cdot \varepsilon_1)(k_1\cdot p_2)(k_2\cdot
p_1)\Big)\Big|_{q = p_1 - k_2 = p_2 -
  k_1}\nonumber\\ &+&\frac{\beta^2}{M^2_{\rm Pl}}\,\frac{1}{q^2 + i
  0}\,\Big( (\varepsilon^*_2\cdot q)(\varepsilon_1\cdot q)(p_1\cdot
p_2) - (\varepsilon^*_2\cdot q)(\varepsilon_1\cdot p_1)(p_2\cdot
q)\nonumber\\ && - (\varepsilon^*_2\cdot p_2)(\varepsilon_1\cdot
q)(p_1\cdot q) + (\varepsilon^*_2\cdot \varepsilon_1)(p_1\cdot
q)(p_2\cdot q)\Big)\Big|_{q = p_1 - k_2 = p_2 -
  k_1}\nonumber\\ &-&\frac{\beta^2}{M^2_{\rm Pl}}\,\frac{1}{q^2 + i
  0}\,\Big((\varepsilon^*_2\cdot p_1)(\varepsilon_1\cdot q)(k_2\cdot
p_2) - (\varepsilon^*_2\cdot p_2)(\varepsilon_1\cdot q)(k_2\cdot
p_1)\nonumber\\ && - (\varepsilon^*_2\cdot p_1)(\varepsilon_1\cdot
k_2)(p_2\cdot q) + (\varepsilon^*_2\cdot \varepsilon_1)(k_2\cdot
p_1)(p_2\cdot q)\Big)\Big|_{q = p_1 - k_2 = p_2 - k_1}
\nonumber\\ &-&\frac{\beta^2}{M^2_{\rm Pl}}\,\frac{1}{q^2 + i
  0}\,\Big((\varepsilon^*_2\cdot q)(\varepsilon_1\cdot p_2)(k_1\cdot
p_1) - (\varepsilon^*_2\cdot k_1)(\varepsilon_1\cdot p_2)(p_1\cdot
q)\nonumber\\ && - (\varepsilon^*_2\cdot q)(\varepsilon_1\cdot
p_1)(k_1\cdot p_2) + (\varepsilon^*_2\cdot \varepsilon_1)(k_1\cdot
p_2)(p_1\cdot q)\Big)\Big|_{q = p_1 - k_2 = p_2 - k_1}.
\end{eqnarray}
The total amplitude of the photon--chameleon scattering
is defined by the sum of the amplitudes Eq.(\ref{eq:40})
\begin{eqnarray}\label{eq:49}
M(\gamma\,\phi \to \phi\,\gamma) = \sum_{j =
  a,b,c,d}M^{(j)}(\gamma\,\phi \to \phi\,\gamma).
\end{eqnarray}
Now let us check gauge invariance of the amplitude of the
photon--chameleon scattering Eq.(\ref{eq:42}). As we have found
already the amplitudes $M^{(a)}(\gamma\,\phi \to \phi\,\gamma)$ and
$M^{(b)}(\gamma\,\phi \to \phi\,\gamma)$ do not depend on the
longitudinal part of the photon propagator, i.e. on the gauge
parameter $\xi$. Then, according to general theory of photon--particle
$(\gamma\,h)$ interactions \cite{Itzykson1980}, the amplitude of
photon--particle scattering should vanish, when the polarisation
vector of the photon either in the initial or in the final state is
replaced by the photon momentum. This means that replacing either
$\varepsilon_1 \to k_1$ or $\varepsilon^*_2 \to k_2$ one has to get
zero for the total amplitude Eq.(\ref{eq:49}). Since one may see that
the amplitude $M^{(d)}(\gamma\,\phi \to \phi\,\gamma)$ is self--gauge
invariant, one has to check the vanishing of the sum of the
amplitudes, defined by the first three Feynman diagrams in
Fig.\,{\ref{fig:torsion2}}, i.e.
\begin{eqnarray}\label{eq:50}
\tilde{M}(\gamma\,\phi \to \phi\,\gamma) = \sum_{j =
  a,b,c}M^{(j)}(\gamma\,\phi \to \phi\,\gamma).
\end{eqnarray}
Replacing $\varepsilon_1 \to k_1$ we obtain
\begin{eqnarray}\label{eq:51}
M^{(a)}(\gamma\,\phi \to \phi\,\gamma)\Big|_{\varepsilon_1 \to k_1}
&=&\frac{\beta^2}{M^2_{\rm Pl}}\Big(- (\varepsilon^*_2\cdot
p_2)(k_1\cdot p_1) + (\varepsilon^*_2\cdot k_1)\,q^2\Big)\Big|_{q =
  p_1 + k_1 = p_2 + k_2},\nonumber\\ M^{(b)}(\gamma\,\phi \to
\phi\,\gamma)\Big|_{\varepsilon_1 \to k_1} &=&\frac{\beta^2}{M^2_{\rm
    Pl}}\Big(- (\varepsilon^*_2\cdot p_1)(k_1\cdot p_2) +
(\varepsilon^*_2\cdot k_1)\,q^2\Big)\Big|_{q = p_1 - k_2 = p_2 -
  k_1},\nonumber\\ M^{(c)}(\gamma\,\phi \to
\phi\,\gamma)\Big|_{\varepsilon_1 \to k_1} &=&\frac{\beta^2}{M^2_{\rm
    Pl}}\Big( - 2(\varepsilon^*_2\cdot k_1)(p_1 \cdot p_2) +
(\varepsilon^*_2\cdot p_2)(k_1\cdot p_1) + (\varepsilon^*_2\cdot
p_1)(k_1\cdot p_2)\Big).
\end{eqnarray}
Because of the relation
\begin{eqnarray}\label{eq:52}
q^2\Big|_{q = p_1 + k_1 = p_2 + k_2} + q^2\Big|_{q = p_1 - k_2 = p_2 -
  k_1} = 2(p_1\cdot p_2)
\end{eqnarray}
the sum of the amplitudes Eq.(\ref{eq:51}) vanishes, i.e.
\begin{eqnarray}\label{eq:53}
\tilde{M}(\gamma\,\phi \to \phi\,\gamma)\Big|_{\varepsilon_1 \to k_1}
= \sum_{j = a,b,c}M^{(j)}(\gamma\,\phi \to
\phi\,\gamma)\Big|_{\varepsilon_1 \to k_1} = 0.
\end{eqnarray}
The same result one may obtain replacing $\varepsilon^*_2 \to k_2$.
Thus, the obtained results confirm gauge invariance of the amplitude
of the photon--chameleon scattering, the complete set of Feynman
diagrams of which is shown in Fig.\,{\ref{fig:torsion2}}.

Of course, because of the smallness of the constant $\beta^4/M^4_{\rm
  Pl} < 10^{-60}\,{\rm barn/eV^2}$, estimated for $\beta < 5.8\times
10^8$ \cite{Jenke2014}, the cross section for the photon--chameleon
scattering is extremely small and hardly plays any important
cosmological role at low energies, for example, for a formation of the
cosmological microwave background and so on
\cite{Davis2009,Lewis2006}. Nevertheless, the observed gauge
invariance of the amplitude of the photon--chameleon scattering is
important for the subsequent extension of the minimal coupling
inclusion of a torsion field to the Weinberg--Salam electroweak model
\cite{Itzykson1980} in the Einstein--Cartan gravity. One of the
interesting consequences of the observed gauge invariance of the
chameleon--photon interaction might be unrenormalisability of the
coupling constant $\beta/M_{\rm Pl}$ by the contributions of all
possible interactions. This might mean that the upper bound on the
chameleon--matter coupling constant $\beta < 5.8 \times 10^8$,
measured in the qBounce experiments with ultracold neutrons
\cite{Jenke2014}, should not be change by taking into account the
contributions of some other possible interactions.

In this connection the results, obtained in this section, can be of
interest with respect to the analysis of the contributions of the
photon--chameleon direct transitions in the magnetic field to the
cosmological microwave background \cite{Davis2009}. The effective
chameleon--photon coupling constant $g_{\rm eff}$, introduced by
Davis, Schelpe and Shaw \cite{Davis2009}, in our approach is equal to
$g_{\rm eff} = \beta/M_{\rm Pl}$. Using the experimental upper bound
$\beta < 5.8\times 10^8$ we obtain $g_{\rm eff} < 2.4\times
10^{-10}\,{\rm GeV^{-1}}$. This constraint is in qualitative agreement
with the results, obtained by Davis, Schelpe and Shaw
\cite{Davis2009}. The experimental constraints on the
chameleon--matter coupling $\beta < 1.9\times 10^7\;(n = 1)$, $\beta <
5.8\times 10^7\; (n = 2)$, $\beta < 2.0 \times 10^8\,(n = 3)$ and
$\beta < 4.8 \times 10^8\,(n = 4)$, measured recently by H. Lemmel
{\it et al.} \cite{Lemmel2015} using the neutron interferometer, place
more strict constraints of the astrophysical sources of chameleons,
investigated in \cite{Brax2007}--\cite{Davis2009}.  

\section{Torsion gravity and Weinberg--Salam electroweak model without
 fermions}
\label{sec:higgs}

In this section we investigate the Weinberg--Salam electroweak model
without fermions in the minimal coupling approach to the torsion field
(see \cite{Hojman1978}), caused by the chameleon field
\cite{Ivanov2014}.

According to Hojman {\it et al.} \cite{Hojman1978}, in the
Einstein--Cartan gravity with a torsion field, induced by a scalar
field, the covariant derivative of a charged (pseudo)scalar particle
with electric charge $q$ should be equal to
\begin{eqnarray}\label{eq:54}
D_{\mu} = \partial_{\mu} - i\,q\,f^{-1}\,A_{\mu}
\end{eqnarray}
with $f = e^{\,\beta\phi/M_{\rm Pl}}$.  Using such a definition of the
covariant derivative one may calculate the electromagnetic field
strength tensor ${\cal F}_{\mu\nu}$ as follows \cite{Itzykson1980}
\begin{eqnarray}\label{eq:55}
{\cal F}_{\mu\nu} = - \frac{1}{iq}\,f\,[D_{\mu},D_{\nu}] = f\,\Big(\partial_{\mu}\Big(f^{-1}\,A_{\nu}\Big) - \partial_{\nu}\Big(f^{-1}\,A_{\mu}\Big)\Big) = A_{\nu,\mu} - A_{\mu,\nu} - A_{\alpha}
{{\cal T}^{\alpha}_{\mu\nu}},
\end{eqnarray}
where the torsion tensor field ${{\cal T}^{\alpha}_{\mu\nu}}$ is given
by Eq.(\ref{eq:11}).  In
this section we discuss the Weinberg--Salam electroweak model
\cite{Itzykson1980} in the Einstein--Cartan gravity with a torsion
field, caused by the chameleon field. Below we consider the
Weinberg--Salam electroweak model without fermions.

The Lagrangian of the Weinberg--Salam electroweak model of the
electroweak bosons and the Higgs field with gauge $SU(2)\times U(1)$
symmetry, determined in the Minkowski space--time, takes the form
\cite{Itzykson1980}
\begin{eqnarray}\label{eq:56}
{\cal L}_{\rm ew} &=& - \frac{1}{4}\,\vec{A}_{\mu\nu}\cdot
\vec{A}^{\mu\nu} - \frac{1}{4}\,B_{\mu\nu}B^{\mu\nu} +
\Big(\partial_{\mu}\Phi - i\,\frac{1}{2}\,g'\,Y_w\,B_{\mu}\Phi -
i\,g\,\vec{I}_w\cdot
\vec{A}_{\mu}\Phi\Big)^{\dagger}\nonumber\\ &&\times\,\Big(\partial^{\mu}\Phi
- i\,\frac{1}{2}\,g''\,Y_w\,B^{\mu}\Phi - i\,g\,\vec{I}_w\cdot
\vec{A}^{\mu}\Phi\Big) - V(\Phi^{\dagger}\Phi),
\end{eqnarray}
where $g'$ and $g$ are the electroweak coupling constants and
$\vec{A}_{\mu}$ and $B_{\mu}$ are vector fields and $\Phi$ is the
Higgs boson field. Then, $Y_w$ and $\vec{I}_w =
\frac{1}{2}\,\vec{\tau}_w$ are the weak hypercharge and the weak
isopin, respectively: $\vec{\tau}_w = (\tau^1_w,\tau^2_w,\tau^3_w)$
are the weak isospin $2\times 2$ Pauli matrices $\tau^a_w \tau^b_w =
\delta^{ab} + i\,\varepsilon^{abc}\tau^c_w$ and ${\rm
  tr}(\tau^a_w\tau^b_w) = 2\,\delta^{ab}$ \cite{Itzykson1980}. The
weak hypercharge $Y_w$ and the third component of the weak isospin
$I_{w3}$ are related by $Q = I_{w3} + Y_w/2$, where $Q$ is the
electric charge of the field in units of the proton electric charge
$e$ \cite{Itzykson1980}. In the Weinberg--Salam electroweak model the
Higgs boson field $\Phi$ possesses the weak isospin $I_w = 1/2$ and the weak
hypercharge $Y_w = 1$. The field strength tensors $\vec{A}_{\mu\nu}$
and $B_{\mu\nu}$ are equal to
\begin{eqnarray}\label{eq:57}
\vec{A}_{\mu\nu} &=& \partial_{\mu}\vec{A}_{\nu} -
\partial_{\nu}\vec{A}_{\mu} + g\,\vec{A}_{\mu} \times
\vec{A}_{\nu},\nonumber\\
B_{\mu\nu} &=& \partial_{\mu}B_{\nu} - \partial_{\nu}B_{\mu}.
\end{eqnarray}
The Higgs boson field $\Phi$ and its vacuum expectation value are
given in the standard form  \cite{Itzykson1980}
\begin{eqnarray}\label{eq:58}
\Phi = \Big(\begin{array}{c}\Phi^+\\ \Phi^0 
\end{array}\Big) \quad,\quad \langle \Phi\rangle = \frac{1}{\sqrt{2}}\,
\Big(\begin{array}{c} 0\\ v 
\end{array}\Big),
\end{eqnarray}
where $\Phi^0 = (v + \varphi)/\sqrt{2}$ and $\varphi$ is a physical
Higgs boson field.  The potential energy density
$V(\Phi^{\dagger}\Phi)$ has also the standard form \cite{Itzykson1980}
\begin{eqnarray}\label{eq:59}
V(\Phi^{\dagger}\Phi) = - \mu^2\Phi^{\dagger}\Phi + \kappa\,
(\Phi^{\dagger}\Phi)^2
\end{eqnarray}
with $\mu^2 > 0$, $\kappa > 0$ and $v^2 = \mu^2/\kappa$. The vacuum
expectation value $v^2$ is related to the Fermi coupling constant
$G_F$ by $\sqrt{2}\,G_F v^2 = 1$, where $G_F = 1.16637(1)\times
10^{-11}\,{\rm MeV^{-2}}$ \cite{PDG2014}. The covariant derivative of
the Higgs field is given by \cite{Itzykson1980}
\begin{eqnarray}\label{eq:60}
D_{\mu} = \partial_{\mu} - i\,\frac{1}{2}\,g'B_{\mu}\Phi -
i\,\frac{1}{2}\,g\,\vec{\tau}\cdot \vec{A}_{\mu}.
\end{eqnarray}
Using the covariant derivative Eq.(\ref{eq:60}) we may calculate the
commutator $[D_{\mu},D_{\nu}]$ and obtain the following expression
\begin{eqnarray}\label{eq:61}
[D_{\mu}, D_{\nu}] = - i\,\frac{1}{2}\,g'(\partial_{\mu} B_{\nu} -
\partial_{\nu}B_{\mu}) - i\,g\,\frac{1}{2}\,\vec{\tau}\cdot
(\partial_{\mu}\vec{A}_{\nu} - \partial_{\nu}\vec{A}_{\mu} +
g\,\vec{A}_{\mu} \times \vec{A}_{\nu}) = -
i\,\frac{1}{2}\,g'\,B_{\mu\nu} - i\,\frac{1}{2}\,g\,\vec{\tau}\cdot
\vec{A}_{\mu\nu},
\end{eqnarray}
where $\vec{A}_{\mu\nu}$ and $B_{\mu\nu}$ are the field strength
tensors Eq.(\ref{eq:57}). Under gauge transformations
\begin{eqnarray}\label{eq:62}
A_{\mu} \to {^{\Omega}}A_{\mu} &=& \Omega A_{\mu}\Omega^{-1} +
\frac{1}{ig}\,\partial_{\mu}\Omega\,\Omega^{-1},\nonumber\\
B_{\mu} \to {^{\Lambda}}B_{\mu} &=& B_{\mu} + \partial_{\mu}\Lambda,
\end{eqnarray}
where $\Omega$ and $\Lambda$ are the gauge matrix and gauge function,
respectively, and $A_{\mu} = \frac{1}{2}\,\vec{\tau}\cdot
\vec{A}_{\mu}$, the field strength tensors $A_{\mu\nu} =
\frac{1}{2}\,\vec{\tau}\cdot \vec{A}_{\mu\nu}$ and $B_{\mu\nu}$
transform as follows \cite{Itzykson1980}
\begin{eqnarray}\label{eq:63}
A_{\mu\nu} \to {^{\Omega}}A_{\mu\nu} &=& \Omega A_{\mu\nu}\Omega^{-1},\nonumber\\
B_{\mu\nu} \to {^{\Lambda}}B_{\mu\nu} &=& B_{\mu\nu}.
\end{eqnarray}
In the Einstein--Cartan gravity with a torsion field in the minimal
coupling approach the covariant derivative Eq.(\ref{eq:60}) should be
taken in the following form
\begin{eqnarray}\label{eq:64}
D_{\mu} = \partial_{\mu} - i\,\frac{1}{2}\,g'\,f^{-1}\,B_{\mu} - i\,\frac{1}{2}\,g\,f^{-1}\,\vec{\tau}\cdot \vec{A}_{\mu}.
\end{eqnarray}
For the definition of field strength tensors $\vec{A}_{\mu\nu}$ and
$B_{\mu\nu}$, extended by the contribution of a torsion field, we
propose to calculate the commutator $[D_{\mu}, D_{\nu}]$. The result
of the calculation is
\begin{eqnarray}\label{eq:65}
[D_{\mu},D_{\nu}] = - i\,\frac{1}{2}\,g'\,f^{-1}\,{\cal B}_{\mu\nu} -
i\,\frac{1}{2}\,g\,f^{-1}\,\vec{\tau}\cdot \vec{{\cal A}}_{\mu\nu},
\end{eqnarray}
where the field strength tensors $\vec{{\cal A}}_{\mu\nu}$ and ${\cal
  B}_{\mu\nu}$ are equal to
\begin{eqnarray}\label{eq:66}
\vec{{\cal A}}_{\mu\nu} &=&\vec{A}_{\nu,\mu} - \vec{A}_{\mu,\nu} +
g\,f^{-1}\,\vec{A}_{\mu} \times \vec{A}_{\nu} -
\vec{A}_{\alpha} {{\cal T}^{\alpha}}_{\mu\nu},\nonumber\\ {\cal
  B}_{\mu\nu} &=& B_{\nu,\mu} - B_{\mu,\nu} - B_{\alpha} {{\cal
    T}^{\alpha}}_{\mu\nu},
\end{eqnarray}
where the torsion tensor field ${{\cal T}^{\alpha}}_{\mu\nu}$ is given
in Eq.(\ref{eq:11}).  Thus, the Lagrangian of electroweak interactions
in the Einstein--Cartan gravity with a torsion field in the minimal
coupling constant approach and the chameleon field, coupled through
the Jordan metric $\tilde{g}_{\mu\nu} = f^2\,g_{\mu\nu}$, takes the
form \cite{Dicke1962}
\begin{eqnarray}\label{eq:67}
\frac{{\cal L}_{\rm ew}}{\sqrt{-g}} &=& - \frac{1}{4}\,\vec{{\cal
    A}}_{\mu\nu}\cdot \vec{{\cal A}}^{\mu\nu} - \frac{1}{4}\,{\cal
  B}_{\mu\nu}{\cal B}^{\mu\nu} + f^2\,\Big(\partial_{\mu}\Phi -
i\,\frac{1}{2}\,g'\,f^{-1}\,B_{\mu}\Phi -
i\,\frac{1}{2}\,g\,f^{-1}\,\vec{\tau}\cdot
\vec{A}_{\mu}\Phi\Big)^{\dagger}\nonumber\\ &&\times\,\Big(\partial^{\mu}\Phi
- i\,\frac{1}{2}\,g'\,f^{-1}\,B^{\mu}\Phi -
i\,\frac{1}{2}\,g\,f^{-1}\,\vec{\tau}\cdot \vec{A}^{\mu}\Phi\Big) -
f^4\,V(\Phi^{\dagger}\Phi),
\end{eqnarray}
where the factor $f^4$ comes from $\sqrt{-\tilde{g}} = f^4\,\sqrt{-
  g}$.  The physical vector boson states of the Weinberg--Salam
electroweak model are \cite{Itzykson1980}
\begin{eqnarray}\label{eq:68}
W^{\pm}_{\mu} &=& \frac{1}{\sqrt{2}}(A^1_{\mu} \mp i A^2_{\mu}),\nonumber\\
Z_{\mu} &=& \sin\theta_W\,B_{\mu} - \cos\theta_W\,A^3_{\mu} ,\nonumber\\
A_{\mu} &=& \cos\theta_W\,B_{\mu} + \sin\theta_W\,A^3_{\mu},
\end{eqnarray}
where $W^{\pm}_{\mu}$ and $Z_{\mu}$ are the electroweak $W$--boson and
$Z$--boson fields and $A_{\mu}$ is the electromagnetic field,
respectively, and $\theta_W$ is the Weinberg angle defined by
$\tan\theta_W = g'/g$.  The electromagnetic coupling constant $e$ as a
function of the coupling constants $g$ and $g'$ is given by $e =
gg'/\sqrt{g^2 + g^{'2}} = g\,\sin\theta_W = g'\cos\theta_W$
\cite{Itzykson1980}. In terms of the electroweak boson fields
$W^{\pm}_{\mu}$ and $Z_{\mu}$, the electromagnetic field $A_{\mu}$,
the Higgs boson field $\varphi$ and the chameleon field $\phi$,
coupled to the gravitational field with torsion, the Lagrangian of the
electroweak interactions takes the following form
\begin{eqnarray}\label{eq:69}
\frac{{\cal L}_{\rm ew}}{\sqrt{- g}} &=& - \frac{1}{4}\,{\cal
  A}_{\mu\nu}{\cal F}^{\mu\nu} - \frac{1}{4}\,{\cal Z}_{\mu\nu}{\cal
  Z}^{\mu\nu} - \frac{1}{2}\,{\cal W}^+_{\mu\nu}{\cal W}^{-\mu\nu} +
\frac{1}{2}\,M^2_W\,W^+_{\mu}W^{-\mu} +
\frac{1}{2}\,M^2_Z\,Z_{\mu}Z^{\mu}\nonumber\\ &&+
\frac{1}{2}\,f^{-1}\,i\,g\,{\cal W}^+_{\mu\nu}\,\Big[\sin\theta_W
  \,\Big(W^{-\mu}A^{\nu} - A^{\mu} W^{-\nu}\Big) - \cos\theta_W
  \,\Big(W^{-\mu}Z^{\nu} - Z^{\mu} W^{-\nu}\Big)\Big]\nonumber\\ &&-
\frac{1}{2}\,f^{-1}\,i\,g\,{\cal W}^-_{\mu\nu}\,\Big[\sin\theta_W
  \,\Big(W^{+\mu}A^{\nu} - A^{\mu} W^{+\nu}\Big) - \cos\theta_W
  \,\Big(W^{+\mu}Z^{\nu} - Z^{\mu} W^{+\nu}\Big)\Big]\nonumber\\ &&-
\frac{1}{2}\,f^{-1}\,i\,g\,\Big(\sin\theta_W\,{\cal A}_{\mu\nu} -
\cos\theta_W\,{\cal Z}_{\mu\nu}\Big)\Big(W^{-\mu}W^{+\nu} -
W^{-\nu}W^{+\mu}\Big)\nonumber\\ &&-
\frac{1}{4}\,f^{-2}\,g^2\,\Big(W^-_{\mu}W^+_{\nu} -
W^-_{\nu}W^+_{\mu}\Big)\Big(W^{+\mu}W^{-\nu} -
W^{+\nu}W^{-\mu}\Big)\nonumber\\ &&-
\frac{1}{2}\,f^{-2}\,g^2\,\Big[\sin\theta_W \,\Big(W^+_{\mu}A_{\nu} -
  A_{\mu} W^+_{\nu}\Big) - \cos\theta_W \,\Big(W^+_{\mu}Z_{\nu} -
  Z_{\mu} W^+_{\nu}\Big)\Big]\nonumber\\ &&\times\,\Big[\sin\theta_W
  \,\Big(W^{-\mu}A^{\nu} - A^{\mu} W^{-\nu}\Big) - \cos\theta_W
  \,\Big(W^{-\mu}Z^{\nu} - Z^{\mu} W^{-\nu}\Big)\Big]\nonumber\\ &&+
\frac{1}{2}\,g\,M_W\,\varphi\,W^+_{\mu}W^{-\mu} +
\frac{1}{8}\,g^2\,\varphi^2\,W^+_{\mu}W^{-\mu} + 
\frac{1}{2}\,\frac{g}{\cos \theta_W}\,M_Z\,\varphi\,Z_{\mu}Z^{\mu} +
\frac{1}{8}\,\frac{g^2}{\cos^2
  \theta_W}\,\varphi^2\,Z_{\mu}Z^{\mu}\nonumber\\ &&+
\frac{1}{2}\,f^2\,\varphi_{,\mu}\varphi^{,\mu} -
\frac{1}{2}\,f^4\,M^2_{\varphi}\,\varphi^2 - f^4\,\kappa\,v\,\varphi^3
- \frac{1}{4}\,f^4\,\kappa\,\varphi^4,
\end{eqnarray}
where $M^2_W = g^2 v^2/4$, $M^2_Z = M^2_W/cos^2\theta_W$ and
$M^2_{\varphi} = 2\,\kappa\,v^2$ are the squared masses of the
$W^{\pm}$--boson, $Z$--boson and Higgs boson field, respectively,
${\cal A}_{\mu\nu}$, ${\cal Z}_{\mu\nu}$ and ${\cal W}^{\pm}_{\mu\nu}$
are the strength field tensors of the electromagnetic, $Z$--boson and
$W^{\pm}$--boson fields. They are equal to
\begin{eqnarray}\label{eq:70}
{\cal A}_{\mu\nu} &=& A_{\nu,\mu} - A_{\mu,\nu} - A_{\sigma}{{\cal
    T}^{\sigma}}_{\mu\nu},\nonumber\\ {\cal Z}_{\mu\nu} &=&
Z_{\nu,\mu} - Z_{\mu,\nu} - Z_{\sigma}{{\cal
    T}^{\sigma}}_{\mu\nu},\nonumber\\ {\cal W}^{\pm}_{\mu\nu} &=&
W^{\pm}_{\nu,\mu} - W^{\pm}_{\mu,\nu} - W^{\pm}_{\sigma}{{\cal
    T}^{\sigma}}_{\mu\nu}.
\end{eqnarray}
Now we are able to extend the obtained results to fermions.

\section{Torsion gravity with chameleon field and Weinberg--Salam 
electroweak model with fermions}
\label{sec:fermion}

\subsection{Dirac fermions with mass $m$ in the Einstein--Cartan 
gravity, coupled to the chameleon field through the Jordan metric 
$\tilde{g}_{\mu\nu} = f^2\,g_{\mu\nu}$}

The Dirac equation in an arbitrary (world) coordinate system is
specified by the metric tensor $g_{\mu\nu}(x)$. It defines an
infinitesimal squared interval between two events
\begin{eqnarray}\label{eq:71}
\hspace{-0.3in}ds^2 = g_{\mu\nu}(x)dx^{\mu}dx^{\nu}.
\end{eqnarray}
The relativistic invariant form of the Dirac equation in an arbitrary
coordinate system is \cite{Fischbach1981}
\begin{eqnarray}\label{eq:72}
\hspace{-0.3in}(i\gamma^{\mu}(x)\nabla_{\mu} - m)\psi(x) = 0,
\end{eqnarray}
where $\gamma^{\mu}(x)$ are a set of Dirac matrices satisfying the
anticommutation relation
\begin{eqnarray}\label{eq:73}
\hspace{-0.3in}\gamma^{\mu}(x)\gamma^{\nu}(x) +
\gamma^{\nu}(x)\gamma^{\mu}(x) = 2 g^{\mu\nu}(x)
\end{eqnarray}
and $\nabla_{\mu}$ is a covariant derivative without gauge fields. For
an exact definition of the Dirac matrices $\gamma^{\mu}(x)$ and the
covariant derivative $\nabla_{\mu}$ we follow \cite{Fischbach1981} and
use a set of tetrad (or vierbein) fields $e^{\hat{\alpha}}_{\mu}(x)$ at each
spacetime point $x$ defined by
\begin{eqnarray}\label{eq:74}
\hspace{-0.3in}dx^{\hat{\alpha}} = e^{\hat{\alpha}}_{\mu}(x) dx^{\mu}.
\end{eqnarray}
The tetrad fields relate in an arbitrary (world) coordinate system a
spacetime point $x$, which is characterised by the index $\mu =
0,1,2,3$, to a locally Minkowskian coordinate system erected at a
spacetime point $x$, which is characterised by the index $\hat{\alpha}
= 0,1,2,3$. The tetrad fields $e^{\hat{\alpha}}_{\mu}(x)$ are related
to the metric tensor $g_{\mu\nu}(x)$ as follows:
\begin{eqnarray}\label{eq:75}
\hspace{-0.3in}ds^2 = \eta_{\hat{\alpha}\hat{\beta}}
\,dx^{\hat{\alpha}} dx^{\hat{\beta}} =
\eta_{\hat{\alpha}\hat{\beta}}\,[e^{\hat{\alpha}}_{\mu}(x)dx^{\mu}]
    [e^{\hat{\beta}}_{\nu}(x)dx^{\nu}] =
    [\eta_{\hat{\alpha}\hat{\beta}}\,e^{\hat{\alpha}}_{\mu}(x)
      e^{\hat{\beta}}_{\nu}(x)] dx^{\mu}dx^{\nu} =
    g_{\mu\nu}(x)dx^{\mu}dx^{\nu}.
\end{eqnarray}
This gives 
\begin{eqnarray}\label{eq:76}
\hspace{-0.3in} g_{\mu\nu}(x) =
\eta_{\hat{\alpha}\hat{\beta}}\,e^{\hat{\alpha}}_{\mu}(x)
e^{\hat{\beta}}_{\nu}(x).
\end{eqnarray}
Thus, the tetrad fields can be viewed as the square root of the metric
tensor $g_{\mu\nu}(x)$ in the sense of a matrix equation
\cite{Fischbach1981}. Inverting the relation Eq.(\ref{eq:74}) we
obtain
\begin{eqnarray}\label{eq:77}
\hspace{-0.3in} \eta_{\hat{\alpha}\hat{\beta}}= g_{\mu\nu}(x)
e^{\mu}_{\hat{\alpha}}(x) e^{\nu}_{\hat{\beta}}(x).
\end{eqnarray}
There are also the following relations
\begin{eqnarray}\label{eq:78}
e^{\mu}_{\hat{\alpha}}(x) e^{\hat{\beta}}_{\mu}(x) &=&
\delta^{\hat{\beta}}_{\hat{\alpha}},\nonumber\\
 e^{\mu}_{\hat{\alpha}}(x)
e^{\hat{\alpha}}_{\nu}(x) &=& \delta^{\mu}_{\nu},\nonumber\\
e^{\mu}_{\hat{\alpha}}(x)e_{\hat{\beta}\mu}(x) &=&
\eta_{\hat{\alpha}\hat{\beta}},\nonumber\\
 e_{\hat{\alpha}\mu}(x) &=&
\eta_{\hat{\alpha}\hat{\beta}}\,e^{\hat{\beta}}_{\mu}(x),\nonumber\\
e^{\hat{\alpha}}_{\mu}(x) e_{\hat{\alpha}\nu}(x) &=& g_{\mu\nu}(x).
\end{eqnarray}
In terms of the tetrad fields $e^{\mu}_{\hat{\alpha}}(x)$ and the
Dirac matrices $\gamma^{\hat{\alpha}}$ in the Minkowski spacetime the
Dirac matrices $\gamma^{\mu}(x)$ are defined by
\begin{eqnarray}\label{eq:79}
\hspace{-0.3in} \gamma^{\mu}(x) = e^{\mu}_{\hat{\alpha}}(x)\gamma^{\hat{\alpha}}.
\end{eqnarray}
A covariant derivative $\nabla_{\mu}$ we define as \cite{Fischbach1981}
\begin{eqnarray}\label{eq:80}
\hspace{-0.3in} \nabla_{\mu} = \partial_{\mu} - \Gamma_{\mu}(x).
\end{eqnarray}
The spinorial affine connection $\Gamma_{\mu}(x)$ is defined by
\cite{Fischbach1981}
\begin{eqnarray}\label{eq:81}
\hspace{-0.3in} \Gamma_{\mu}(x) &=&
\frac{i}{4}\,\sigma^{\hat{\alpha}\hat{\beta}}\, e^{\nu}_{\hat{\alpha}}(x)
e_{\hat{\beta}\nu;\mu}(x),
\end{eqnarray}
where $\sigma^{\hat{\alpha}\hat{\beta}} = {\textstyle
  \frac{i}{2}}(\gamma^{\hat{\alpha}}\gamma^{\hat{\beta}} -
\gamma^{\hat{\beta}}\gamma^{\hat{\alpha}})$ and
$e_{\hat{\beta}\nu;\mu}(x)$ is given in terms of the affine connection
$\Gamma^{\alpha}_{\mu\nu}(x)$
\begin{eqnarray}\label{eq:82}
\hspace{-0.3in} e_{\hat{\beta}\nu;\mu}(x) &=& e_{\hat{\beta}\nu,\mu}(x) -
{\Gamma^{\alpha}}_{\mu\nu}(x)\, e_{\hat{\beta}\alpha}(x),\nonumber\\
 e^{\hat{\beta}}_{\nu;\mu}(x) &=& e^{\hat{\beta}}_{\nu,\mu}(x) -
{\Gamma^{\alpha}}_{\mu\nu}(x)\, e^{\hat{\beta}}_{\alpha}(x).
\end{eqnarray}
In the Einstein gravity the affine connection
${\Gamma^{\alpha}}_{\mu\nu}(x)$ is equal to ${\Gamma^{\alpha}}_{\mu\nu}(x)
= \{{^{\alpha}}_{\mu\nu}\}$ (see Eq.(\ref{eq:10})). Specifying the spacetime metric one may transform the Dirac equation Eq.(\ref{eq:72}) into the standard form
\begin{eqnarray}\label{eq:83}
\hspace{-0.3in} i\frac{\partial \psi(t,\vec{r}\,)}{\partial t}
=\hat{\rm H}\,\psi(t,\vec{r}\,),
\end{eqnarray}
where $\hat{\rm H}$ is the Hamilton operator. For example, for the
static metric $ds^2 = V^2\,dt^2 - W^2\,(d\vec{r}\,)^2$, where $V$ and
$W$ are spatial functions, one may show that the Hamilton operator is
given by (see \cite{Obukhov2001,Ivanov2014})
\begin{eqnarray}\label{eq:84}
  \hat{\rm H} = \gamma^0mV -
  i\,\frac{V}{W}\,\gamma^0\vec{\gamma}\cdot\Big(\vec{\nabla} +
  \frac{\vec{(\nabla}V)}{2 V} + \frac{\vec{(\nabla}W)}{W}\Big).
\end{eqnarray}
In the approach to the Einstein--Cartan gravity with torsion,
developed above, the Lagrangian of the Dirac field $\psi(x)$ with mass
$m$, coupled to the chameleon field through the Jordan metric
$\tilde{g}_{\mu\nu} = f^2\,g_{\mu\nu}$, is equal to
\begin{eqnarray}\label{eq:85}
  {\cal L}_m &=&
  \sqrt{-\tilde{g}}\,\bar{\psi}\Big(i\tilde{g}^{\mu\nu}\tilde{\gamma}_{\mu}(x)\tilde{\nabla}_{\nu}
  - m\Big)\psi = \sqrt{-\tilde{g}}\,\bar{\psi}\Big(i
  \tilde{e}^{\mu}_{\hat{\alpha}}\gamma^{\hat{\alpha}}\tilde{\nabla}_{\mu}
  - m\Big)\psi,
\end{eqnarray}
where the tetrad fields in the Jordan--frame and the Einstein--frame
are related by
\begin{eqnarray}\label{eq:86}
  \tilde{e}^{\hat{\alpha}}_{\mu} &=& f\,e^{\hat{\alpha}}_{\mu},
      \nonumber\\ \tilde{e}^{\mu}_{\hat{\alpha}} &=&
        f^{-1}\,e^{\mu}_{\hat{\alpha}}
\end{eqnarray}
and the covariant derivative is equal to
\begin{eqnarray}\label{eq:87}
  \tilde{\nabla}_{\mu} &=& \nabla_{\mu} = \partial_{\mu} -
  \Gamma_{\mu}(x),\nonumber\\ \Gamma_{\mu}(x)
  &=&\frac{i}{4}\,\sigma^{\hat{\alpha}\hat{\beta}}
  e^{\nu}_{\hat{\alpha}}(x)
  e_{\hat{\beta}\nu;\mu},\nonumber\\ {\tilde{\Gamma}^{\alpha}}_{\mu\nu} &=&
  \{{^{\alpha}}_{\mu\nu}\} + {\delta^{\alpha}}_{\mu}\,f^{-1}\,f_{,\nu} = \{{^{\alpha}}_{\mu\nu}\} + \frac{\beta}{M_{\rm Pl}}\,{\delta^{\alpha}}_{\mu}\,\phi_{,\nu}.
\end{eqnarray}
As a result we get
\begin{eqnarray}\label{eq:88}
  {\cal L}_m &=& \sqrt{-
    g}\,f^4\,\bar{\psi}\Big(f^{-1}i\gamma^{\mu}(x)\nabla_{\mu} -
  m\Big)\psi.
\end{eqnarray}
Below we apply the obtained results to the analysis of the electroweak
interactions of the neutron and proton.

\subsection{Electroweak model for neutron and proton, coupled to 
chameleon field through the Jordan metric $\tilde{g}_{\mu\nu} = f^2\,g_{\mu\nu}$}

For the subsequent application of the results, obtained below, to the
analysis of the contribution of the chameleon field to the radii of
the neutron and the proton and to the neutron $\beta^-$--decay we
defined the electroweak model for the following multiplets
\begin{eqnarray}\label{eq:89}
  N_L &=& \Big(\begin{array}{c} p_L \\ n_L 
\end{array}\Big)\;,\; p_R \;,\; n_R,\nonumber\\
{\ell}_L &=& \Big(\begin{array}{c} \nu_{e L}\\ e^-_L 
\end{array}\Big)\;,\; e^-_R,
\end{eqnarray}
where $\psi_L = \frac{1}{2}(1 - \gamma^5)\,\psi$ and $\psi_R =
\frac{1}{2}(1 + \gamma^5)\,\psi$. Such a model is renormalisable also
because of the vanishing of the contribution of the
Adler--Bell--Jackiw anomalies $Q_p + Q_e = 0$, where $Q_p = +1$ and
$Q_e = -1$ are the electric charges of the proton and electron,
measured in the units of the proton charge $e$ \cite{Jackiw1972}.

The fermion states Eq.(\ref{eq:89}) have the following electroweak
quantum numbers: $N_L: (I_w = \frac{1}{2}, Y_w = 1)$, $p_R :(I_w = 0,
Y_w = 2)$, $n_R: (I_w = 0, Y_w = 0)$ and $\ell_L: (I_w = \frac{1}{2},
Y_w = - 1)$, $e^-_R:( I_w = 0, Y_w = - 2)$, where the third component
of the weak isospin $I_{w 3}$ and weak hypercharge $Y_w$ are related
by $Q = I_{w3} + Y_w/2$. In the Einstein--Cartan gravity with the
torsion field and the chameleon field, coupled to matter field through
the Jordan metric $\tilde{g}_{\mu\nu} = f^2\,g_{\mu\nu}$, the
Lagrangian of the fermion fields Eq.(\ref{eq:89}), coupled to the
vector electroweak boson fields and the Higgs boson field, is equal to
\begin{eqnarray}\label{eq:90}
  \frac{{\cal L}_{\rm ewf}}{\sqrt{- g}} &=&
  f^4\,\bar{N}_L\Big[f^{-1}\,i\gamma^{\mu}(x)\Big(\partial_{\mu} -
    i\,\frac{1}{2}\,g'\,f^{-1}\,B_{\mu} -
    i\,\frac{1}{2}\,g\,f^{-1}\,\vec{\tau}\cdot \vec{A}_{\mu} -
    \Gamma_{\mu}\Big)\Big]\,N_L\nonumber\\ &+&f^4\,\bar{p}_R\Big[f^{-1}\,
    i\gamma^{\mu}(x)\Big(\partial_{\mu} - i\,g'\,f^{-1}\,B_{\mu} -
    \Gamma_{\mu}\Big)\Big]\,p_R +
  f^4\,\bar{n}_R\Big[f^{-1}\,i\gamma^{\mu}(x)\Big(\partial_{\mu} -
    \Gamma_{\mu}\Big)\Big]\,n_R\nonumber\\ &+&
  f^4\,\bar{\ell}_L\Big[f^{-1}\,i\gamma^{\mu}(x)\Big(\partial_{\mu} +
    i\,\frac{1}{2}\,g'\,f^{-1}\,B_{\mu} -
    i\,\frac{1}{2}\,g\,f^{-1}\,\vec{\tau}\cdot \vec{A}_{\mu} -
    \Gamma_{\mu}\Big)\Big]\,{\ell}_L\nonumber\\ &+&
  f^4\,\bar{e}^-_R\Big[f^{-1}\,i\gamma^{\mu}(x)\Big(\partial_{\mu} +
    i\,g'\,f^{-1}\,B_{\mu} - \Gamma_{\mu}\Big)\Big]\,e^-_R.
\end{eqnarray}
The masses of the neutron and electron one may gain by virtue of the
following interactions with the Higgs field $\Phi$
\begin{eqnarray}\label{eq:91}
\frac{ \delta {\cal L}_{\rm hne}}{\sqrt{- g}} &=& - f^4\,\kappa_n\Big(\bar{N}_L
\Phi\,n_R + \bar{n}_R \Phi^{\dagger}N_L) - f^4\,\kappa_e\Big(\bar{\ell}_L
\Phi\,e^-_R + \bar{e}^-_R \Phi^{\dagger}{\ell}_L) =\nonumber\\ &=&
- f^4\,m_n\,\bar{n}n - f^4\,\frac{\kappa_n}{\sqrt{2}}\,\bar{n}n\,\varphi -
f^4\,m_e\,\bar{e}^-e^- -
f^4\,\frac{\kappa_e}{\sqrt{2}}\,\bar{e}^-e^-\,\varphi,
\end{eqnarray}
where $\kappa_n$ and $\kappa_e$ are the input parameters, defining the
neutron and electron masses $m_n = \kappa_nv/\sqrt{2}$ and $m_e =
\kappa_ev/\sqrt{2}$, respectively. In principle, the proton mass we
may gain due to the interaction of the proton fields with the Higgs
field $\bar{\Phi}$, defined by the column matrix with the components
$(v + \varphi)/\sqrt{2}$ and $\Phi^- = (\Phi^+)^{\dagger}$. Using the
Higgs field $\bar{\Phi}$ one gets
\begin{eqnarray}\label{eq:92}
\frac{ \delta {\cal L}_{\rm hp}}{\sqrt{- g}} = - f^4\,\kappa_p\Big(\bar{N}_L
\bar{\Phi}\,p_R + \bar{p}_R \bar{\Phi}^{\dagger}N_L) = -
f^4\,m_p\,\bar{p}p -
f^4\,\frac{\kappa_p}{\sqrt{2}}\,\bar{p}p\,\varphi,
\end{eqnarray}
where $m_p = \kappa_pv/\sqrt{2}$ is the proton mass. In terms of the
physical vector field states the Lagrangian of fermion field is given
by
\begin{eqnarray}\label{eq:93}
  \frac{{\cal L}_{\rm ew}}{\sqrt{- g}} &=&
  f^3\,\bar{p}(x)\,i\gamma^{\mu}(x)\Big(\partial_{\mu} -
  i\,f^{-1}\,e\,A_{\mu}(x) + i\,f^{-1}\,\frac{g}{2\cos\theta_W}(1 -
  2\sin^2\theta_W) Z_{\mu}(x) - \Gamma_{\mu}(x)\Big)\,\Big(\frac{1 -
    \gamma^5}{2}\Big)\,p(x)\nonumber\\ &+&f^3\,\bar{p}(x)
  i\gamma^{\mu}(x)\Big(\partial_{\mu} - i\,f^{-1}\,e\,A_{\mu}(x) +
  i\,f^{-1}\,\frac{g}{\cos\theta_W}\,\sin^2\theta_W\, Z_{\mu}(x) -
  \Gamma_{\mu}(x)\Big)\,\Big(\frac{1 +
    \gamma^5}{2}\Big)\,p(x)\nonumber\\ &+&
  f^3\,\bar{n}(x)\,i\gamma^{\mu}(x)\Big(\partial_{\mu} -
  i\,f^{-1}\,\frac{g}{2\cos\theta_W} Z_{\mu}(x)-
  \Gamma_{\mu}(x)\Big)\,\Big(\frac{1 -
    \gamma^5}{2}\Big)\,n(x)\nonumber\\ &+&
  f^3\,\bar{n}(x)\,i\gamma^{\mu}(x)\Big(\partial_{\mu} -
  \Gamma_{\mu}(x)\Big)\,\Big(\frac{1 +
    \gamma^5}{2}\Big)\,n(x)\nonumber\\ &+&
  f^3\,\frac{g}{\sqrt{2}}\,\bar{p}(x)\gamma^{\mu}(x)\,\Big(\frac{1 -
    \gamma^5}{2}\Big)\,n(x)\,W^+_{\mu}(x) +
  f^3\,\frac{g}{\sqrt{2}}\,\bar{n}(x)\gamma^{\mu}(x)\,\Big(\frac{1 -
    \gamma^5}{2}\Big)\,p(x)\,W^-_{\mu}(x)\nonumber\\ &+&
  f^3\,\bar{\nu}_e(x)\,i\gamma^{\mu}(x)\Big(\partial_{\mu} +
  i\,f^{-1}\,\frac{g}{2\cos\theta_W} Z_{\mu}(x)-
  \Gamma_{\mu}(x)\Big)\,\Big(\frac{1 -
    \gamma^5}{2}\Big)\,\nu_e(x)\nonumber\\ &+&f^3\,\bar{e}^-(x)\,i\gamma^{\mu}(x)\Big(\partial_{\mu}
  + i\,f^{-1}\,e\,A_{\mu}(x) - i\,f^{-1}\,\frac{g}{2\cos\theta_W}(1 -
  2\sin^2\theta_W) Z_{\mu}(x) - \Gamma_{\mu}(x)\Big)\,\Big(\frac{1 -
    \gamma^5}{2}\Big)\,e^-(x)\nonumber\\ &+&f^3\,\bar{e}^-(x)\,i\gamma^{\mu}(x)\Big(\partial_{\mu}
  + i\,f^{-1}\,e\,A_{\mu}(x) +
  i\,f^{-1}\,\frac{g}{\cos\theta_W}\,\sin^2\theta_W) Z_{\mu}(x) -
  \Gamma_{\mu}(x)\Big)\,\Big(\frac{1 +
    \gamma^5}{2}\Big)\,e^-(x)\nonumber\\ &-& f^4\,m_p\,\bar{p}(x)p(x)
  - f^4\,m_n\,\bar{n}(x)n (x)-
  f^4\,m_e\,\bar{e}^-(x)e^-(x)\nonumber\\ &-&
  f^4\,\frac{\kappa_p}{\sqrt{2}}\,\bar{p}(x)p(x)\,\varphi(x) -
  f^4\,\frac{\kappa_n}{\sqrt{2}}\,\bar{n}(x)n(x)\,\varphi(x) -
  f^4\,\frac{\kappa_e}{\sqrt{2}}\,\bar{e}^-(x)e^-(x)\,\varphi(x).
\end{eqnarray}
We note that for the calculation of $\Gamma_{\mu}(x)$ we have to use
the affine connection, given by Eq.(\ref{eq:87}).

\section{Contribution of chameleon field to charge radii of neutron 
and proton and to neutron $\beta^-$--decay}
\label{sec:beta}

The electroweak model in the Einstein--Cartan gravity with the torsion
and chameleon fields, analysed above, is applied to some specific
processes of electromagnetic and weak interactions of the neutron and
proton to the chameleon field. In this section we calculate the
amplitudes of the electron--neutron and electron--proton low--energy
scattering with the chameleon field exchange and define the
contributions of the chameleon field to the charge radii of the
neutron and proton. We calculate also the contribution of the
chameleon field to the energy spectra of the neutron $\beta^-$--decay
and the lifetime of the neutron. The calculations we carry out in the
Minkowski spacetime replacing metric tensor $g_{\mu\nu}$ in the
Einstein frame by the metric tensor of the Minkowski spacetime
$g_{\mu\nu} \to \eta_{\mu\nu}$. The Lagrangian of the electromagnetic
and electroweak interactions of the neutron, the proton, the electron
and the electron neutrino, coupled to the torsion field and the
chameleon field in the Minkowski spacetime, is given by
\begin{eqnarray*}
  {\cal L}_{\rm ew} &=&
  f^3\,\bar{p}(x)\,\Big\{i\gamma^{\mu}\Big(\partial_{\mu} -
  i\,f^{-1}\,e\,A_{\mu}(x) + i\,f^{-1}\,\frac{g}{4\cos\theta_W}\,(1 -
  (1 - 4\sin^2\theta_W)\gamma^5) Z_{\mu}(x) - \Gamma_{\mu}(x)\Big) -
  m_p\,f\Big\}\,p(x)\nonumber\\ &+&
  f^3\,\bar{n}(x)\,\Big\{i\gamma^{\mu} \Big(\partial_{\mu} -
  i\,f^{-1}\,\frac{g}{4\cos\theta_W}(1 - \gamma^5)\, Z_{\mu}(x) -
  \Gamma_{\mu}(x)\Big) - m_n\,f\,\Big\}\,n(x)\nonumber\\ &+&
  f^3\,\bar{\nu}_e(x)\,i\gamma^{\mu}\Big(\partial_{\mu} +
  i\,f^{-1}\,\frac{g}{2\cos\theta_W} Z_{\mu}(x)-
  \Gamma_{\mu}(x)\Big)\,\Big(\frac{1 -
    \gamma^5}{2}\Big)\,\nu_e(x)\nonumber\\ &+&f^3\,\bar{e}^-(x)\,\Big\{i\gamma^{\mu}\Big(\partial_{\mu}
  + i\,f^{-1}\,e\,A_{\mu}(x) - i\,f^{-1}\,\frac{g}{4\cos\theta_W}\,((1
  - 4\sin^2\theta_W) - \gamma^5) Z_{\mu}(x) - \Gamma_{\mu}(x)\Big) -
  m_e\,f\Big\}\,e^-(x)\nonumber\\ 
&+&
  f^3\,\frac{g}{\sqrt{2}}\,\bar{p}(x)\gamma^{\mu}\,\Big(\frac{1 -
    \gamma^5}{2}\Big)\,n(x)\,W^+_{\mu}(x) +
  f^3\,\frac{g}{\sqrt{2}}\,\bar{n}(x)\gamma^{\mu}\,\Big(\frac{1 -
    \gamma^5}{2}\Big)\,p(x)\,W^-_{\mu}(x)\nonumber\\ 
\end{eqnarray*}
\begin{eqnarray}\label{eq:94}
&+&
  f^3\,\frac{g}{\sqrt{2}}\,\bar{\nu}_e(x)\gamma^{\mu}\,\Big(\frac{1 -
    \gamma^5}{2}\Big)\,e^-(x)\,W^+_{\mu}(x) +
  f^3\,\frac{g}{\sqrt{2}}\,\bar{e}^-(x)\gamma^{\mu}\,\Big(\frac{1 -
    \gamma^5}{2}\Big)\,\nu_e(x)\,W^-_{\mu}(x) + \ldots,
\end{eqnarray}
where the ellipsis denotes the interactions with the Higgs
field. Then. $m_p = 938.27205\,{\rm MeV}$, $m_n = 939.56538\,{\rm
  MeV}$ and $m_e = 0.51100\,{\rm MeV}$ are the masses of the proton,
neutron and electron, respectively \cite{PDG2014}. Expanding the
conformal factor $f = e^{\,\beta\phi/M_{\rm Pl}}$ in powers of the
chameleon field and keeping only the linear terms we arrive at the
following interactions
\begin{eqnarray}\label{eq:95}
  {\cal L}_{\rm int} &=& - \beta\,\frac{m_p}{M_{\rm
      Pl}}\,\bar{p}(x)p(x)\,\phi(x) - \beta\,\frac{m_n}{M_{\rm
      Pl}}\,\bar{n}(x)n(x)\,\phi(x) - \beta\,\frac{m_e}{M_{\rm
      Pl}}\,\bar{e}^-(x)e^-(x)\,\phi(x)\nonumber\\ &+&
  e\,\bar{p}(x)\gamma^{\mu}p(x)\,A_{\mu}(x) -
  \frac{g}{4\cos\theta_W}\,\bar{p}(x) \gamma^{\mu}\Big(1 - (1 -
  4\sin^2\theta_W)\gamma^5\Big)p(x) Z_{\mu}(x)\nonumber\\ &-&
  e\,\bar{e}^-(x)\gamma^{\mu}e^-(x)\,A_{\mu}(x) +
  \frac{g}{4\cos\theta_W}\,\bar{p}(x) \gamma^{\mu}\Big((1 -
  4\sin^2\theta_W) - \gamma^5\Big)e^-(x)
  Z_{\mu}(x)\nonumber\\ &+&\frac{g}{4\cos\theta_W}\,\bar{n}(x)
  \gamma^{\mu}\Big(1 - \gamma^5\Big)n(x) Z_{\mu}(x) -
  \frac{g}{4\cos\theta_W}\,\bar{\nu}_e(x)\gamma^{\mu}(1 -
  \gamma^5)\nu_e(x)\, Z_{\mu}(x)\nonumber\\ &+&
  \frac{g}{\sqrt{2}}\,\bar{p}(x)\gamma^{\mu}\,\Big(\frac{1 -
    \gamma^5}{2}\Big)\,n(x)\,W^+_{\mu}(x) +
  \frac{g}{\sqrt{2}}\,\bar{n}(x)\gamma^{\mu}\,\Big(\frac{1 -
    \gamma^5}{2}\Big)\,p(x)\,W^-_{\mu}(x)\nonumber\\ &+&
  \frac{g}{\sqrt{2}}\,\bar{\nu}_e(x)\gamma^{\mu}\,\Big(\frac{1 -
    \gamma^5}{2}\Big)\,e^-(x)\,W^+_{\mu}(x) +
  \frac{g}{\sqrt{2}}\,\bar{e}^-(x)\gamma^{\mu}\,\Big(\frac{1 -
    \gamma^5}{2}\Big)\,\nu_e(x)\,W^-_{\mu}(x),
\end{eqnarray}
where we have omitted the interactions with the Higgs field, which do
not contribute to the processes our interest. The contribution of the
terms, containing $\Gamma_{\mu}(x)$, which in the Minkowski spacetime
is equal to $\Gamma_{\mu} = \frac{i}{4}\,\sigma_{\mu\nu}({\ell
  n}f)^{,\nu}$, can be transformed into total divergences and omitted.

\subsection{Contributions of chameleon  to squared 
charge radius of neutron $r^2_n$}

\begin{figure}
\centering
\includegraphics[width=0.35\linewidth]{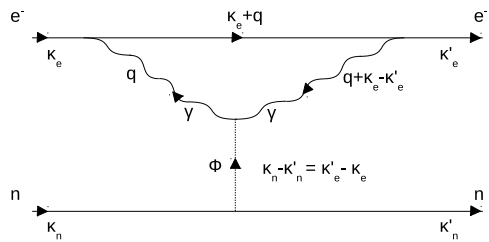}
\caption{Feynman diagram for the contribution of the
  torsion--chameleon field to the squared charge radius of the
  neutron}
 \label{fig:torsion1}
\end{figure}

The torsion (chameleon) contribution to the squared charge radius of
the neutron is defined by the Feynman diagram in
Fig.\,\ref{fig:torsion1}.
The chameleon--neutron interaction is defined by (see
Eq.(\ref{eq:95}))
\begin{eqnarray}\label{eq:96}
{\cal L}_{nn\phi}(x) = -\beta\,\frac{m_n}{M_{\rm
      Pl}}\,\bar{n}(x)n(x)\,\phi(x).
\end{eqnarray}
The Lagrangian of the q$\phi\gamma\gamma$ interaction, given by
Eq.(\ref{eq:30}) for $\sqrt{- g} = 1$, can be transcribed into the
form
\begin{eqnarray}\label{eq:97}
{\cal L}_{\phi \gamma\gamma}(x) = - \frac{1}{2}\,\frac{\beta}{M_{\rm
    Pl}}\,F^{\mu\nu}(x) F_{\mu\nu}(x)\phi(x),
\end{eqnarray}
where we have omitted the total divergence.  The analytical expression
for the Feynman diagram in Fig.\,\ref{fig:torsion1} is equal to
\begin{eqnarray}\label{eq:98}
\hspace{-0.3in}&&M(e^-\,n \to e^-\,n) = e^2\,\frac{\beta^2
  m_n}{M^2_{\rm Pl}}\int \frac{d^4q}{(2\pi)^4
  i}\,\Big[\bar{u}(k'_e)\gamma^{\mu}\frac{1}{m_e - \hat{k}_e - \hat{q} -
  i0}\gamma^{\nu}u(k_e)\Big]\,(\eta^{\sigma\lambda} q^{\varphi} -
\eta^{\sigma\varphi} q^{\lambda})\,D_{\sigma\nu}(q)\nonumber\\
\hspace{-0.3in}&&\times\,(\eta_{\alpha\lambda}(q + k_e - k'_e)_{\varphi} -
\eta_{\alpha\varphi}(q + k_e - k'_e)_{\lambda}) {D^{\alpha}}_{\mu}(q + k_e - k'_e)\,\frac{1}{m^2_{\phi} - (k_n -
  k'_n)^2}\,[\bar{u}(k'_n)u(k_n)],
\end{eqnarray}
where $m_{\phi}$ is the chameleon mass Eq.(\ref{eq:39}) as a function
of the chameleon--matter coupling constant $\beta$, the environment
density $\rho$ and the Ratra--Peebles index $n$, and
$D_{\alpha\beta}(Q)$ is the photon propagator Eq.(\ref{eq:46}).
Substituting the photon propagators $D^{\alpha\mu}(q)$ and
$D_{\sigma\nu}(q + k_e - k'_e)$, taken in the form of
Eq.(\ref{eq:46}), into Eq.(\ref{eq:98}) one may show that the
integrand does not depend on the gauge parameter $\xi$, i.e. the
integrand is gauge invariant.

Measuring the electric charge on the neutron in the electric charge of
the proton for the calculation of the neutron electric radius we have
to compare Eq.(\ref{eq:98}) to the amplitude
\begin{eqnarray}\label{eq:99}
M(e^-n\to e^-n) = - e_qe_p\frac{1}{6}
r^2_n\,[\bar{u}(k'_e)u(k_e)][\bar{u}(k'_n)u(k_n)],
\end{eqnarray}
where $e_q = - e_p$ and $e_p$ are the electric charges of the electron
and proton, respectively, and $r^2_n$ is the squared charge radius of
the neutron. The product $[\bar{u}(k'_e)u(k_e)][\bar{u}(k'_n)u(k_n)]$
is equivalent to the product $[\bar{u}(k'_e) \gamma^0
  u(k_e)][\bar{u}(k'_n) \gamma^0 u(k_n)]$ in the low--energy limit.

\begin{figure}
\centering
\includegraphics[width=0.4\linewidth]{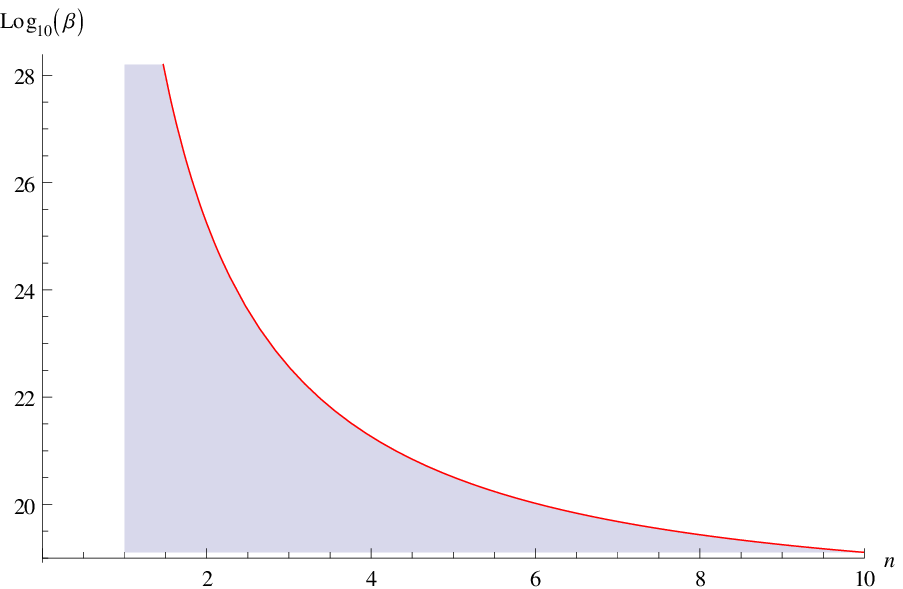}
\includegraphics[width=0.4\linewidth]{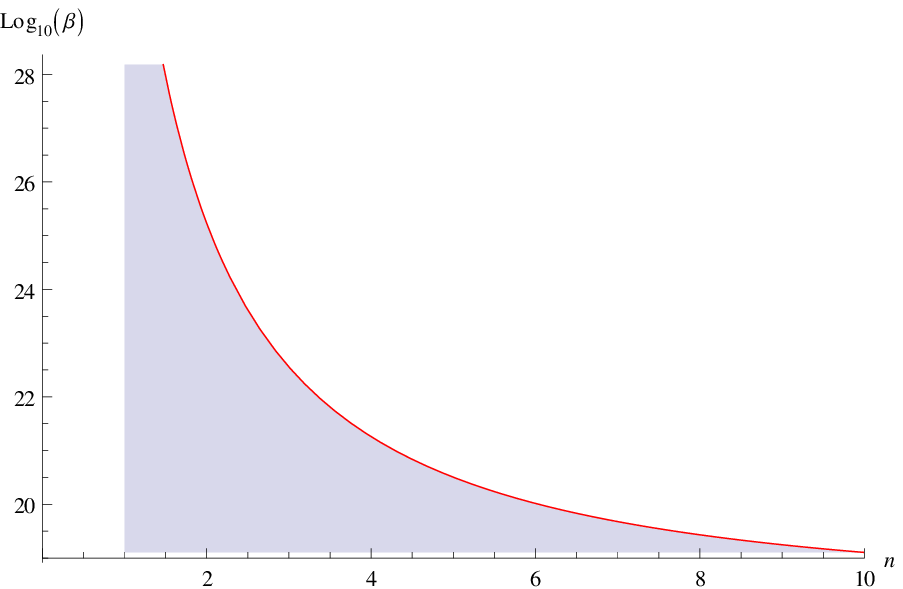}
\caption{The lower bound of the chameleon--matter coupling constant
  $\beta$ from the experiments on the electron--neutron scattering
  length for the scattering of the slow neutron by lead (left) and
  bismuth (right), respectively. The shaded area is excluded.}
 \label{fig:beta}
\end{figure}

From the comparison of Eq.(\ref{eq:98}) with Eq.(\ref{eq:99}) the
contribution of the chameleon to the squared charge radius of the
neutron can be determined by the following analytical expression
\begin{eqnarray}\label{eq:100}
\hspace{-0.3in}&&r^2_n = \frac{6\beta^2 m}{m_e m^2_{\phi} M^2_{\rm
    Pl}}\int \frac{d^4q}{(2\pi)^4
  i}\,\Big[\bar{u}(k_e)\gamma^{\mu}\frac{1}{m_e - \hat{k}_e - \hat{q}
    - i0}\gamma^{\nu}u(k_e)\Big]\,(\eta^{\mu\nu} q^2 -
q^{\mu}q^{\nu})\,\frac{1}{(q^2 + i0)^2},
\end{eqnarray}
where we have set $k'_e = k_e$ and $k'_n = k_n$. Merging denominators
by using the Feynman formula
\begin{eqnarray}\label{eq:101}
\hspace{-0.3in} \frac{1}{A^2 B} = \int^1_0 \frac{2x dx}{[Ax + B(1 -
    x)]^3}
\end{eqnarray}
we arrive at the following expression
\begin{eqnarray}\label{eq:102}
\hspace{-0.3in}&&r^2_n = \frac{6\beta^2 m}{m_e m^2_{\phi} M^2_{\rm
    Pl}}\int^1_0 dx\,2x\int \frac{d^4q}{(2\pi)^4
  i}\,\frac{\bar{u}(k_e)\gamma^{\mu}(m_e + \hat{k}_e +
  \hat{q})\gamma^{\nu}u(k_e)}{[m^2_e(1 - x)^2 - (q + k_e(1 -
  x))^2]^3}\,(\eta^{\mu\nu} q^2 - q^{\mu}q^{\nu}).
\end{eqnarray}
Making use a standard procedure for the calculation of the integrals
Eq.(\ref{eq:102}), i.e. i) the shift of the virtual momentum $q +
k_e(1 - x)$, ii) the integration over the 4--dimensional solid angle
and iii) the Wick rotation, we arrive at the expression
\begin{eqnarray}\label{eq:103}
\hspace{-0.3in}&&r^2_n = - \frac{6\beta^2 m _e m}{m^2_{\phi} M^2_{\rm
    Pl}}\int^1_0 dx\,2x\int \frac{d^4q}{(2\pi)^4}\frac{9(1 - x)q^2 -
  3m^2_e (1 - x)^3}{[q^2 + m^2_e(1 - x)^2]^3} = -
\frac{9}{4\pi^2}\,\frac{\beta^2 m _e m}{ m^2_{\phi} M^2_{\rm
    Pl}}\,{\ell n}\Big(\frac{M}{m_e}\Big),
\end{eqnarray}
where $M$ is the ultra--violet cut--off. For numerical estimates we
set $M = M_{\rm Pl}$ \cite{Feynman1995}. This gives
\begin{eqnarray}\label{eq:104}
\hspace{-0.3in}r^2_n = -
\frac{9}{4\pi^2}\,\frac{\beta^2}{m^2_{\phi}}\,\frac{m _e m}{M^2_{\rm
    Pl}}{\ell n}\Big(\frac{M_{\rm Pl}}{m_e}\Big).
\end{eqnarray}
According to \cite{Kopecky1997}, the squared charge radius of the
neutron can be defined by the expression
\begin{eqnarray}\label{eq:105}
\hspace{-0.3in}r^2_n = \frac{3 b_{ne}}{m \alpha},
\end{eqnarray}
where $\alpha = 1/137.036$ and $b_{ne}$ are the fine--structure
constant and the electron--neutron scattering length,
respectively. For the experimental values of the electron--neutron
scattering lengths $b_{ne} = (-1.330 \pm 0.027 \pm 0.030)\times
10^{-3}\,{\rm fm}$ and $b_{ne} = (-1.440 \pm 0.033 \pm 0.030)\times
10^{-3}\,{\rm fm}$, measured from the scattering of low--energy
electrons by ${^{208}}{\rm Pb}$ and ${^{208}}{\rm Bi}$
\cite{Kopecky1997}, respectively, we get $(r^2_n)_{\exp} = -
0.115(4)\,{\rm fm^2}$ and $(r^2_n)_{\exp} = - 0.124(4)\,{\rm fm^2}$
\cite{Kopecky1997}, respectively. The theoretical value of the squared
charge radius of the neutron is
\begin{eqnarray}\label{eq:106}
\hspace{-0.3in}r^2_n = - 3.588\times
10^{-17}\,\frac{\beta^2}{m^2_{\phi}}\,{\rm fm^2} = - 6.229 \times
10^{-18}\,\frac{\displaystyle \beta^{\textstyle \frac{n}{n + 1}}}{n(n
  + 1)}\,\Big(\frac{n M_{\rm Pl}\Lambda^3}{\rho_m}\Big)^{\textstyle
  \frac{n + 2}{n + 1}}\,{\rm fm^2},
\end{eqnarray}
where the chameleon mass is measured in ${\rm meV}$. From the
comparison to the experimental values we obtain
\begin{eqnarray}\label{eq:107}
\hspace{-0.3in}&&\beta \ge  [1.864 \times 10^{16} n(n + 1)]^{\textstyle
  \frac{n + 1}{n}} \Big(\frac{\rho_m}{n M_{\rm
    Pl}\Lambda^3}\Big)^{\textstyle \frac{n + 2}{n}}\quad {\rm for}\quad
       {^{208}}{\rm Pb},\nonumber\\
\hspace{-0.3in}&&\beta \ge  [1.991 \times 10^{16} n(n + 1)]^{\textstyle
  \frac{n + 1}{n}} \Big(\frac{\rho_m}{n M_{\rm
    Pl}\Lambda^3}\Big)^{\textstyle \frac{n + 2}{n}}\quad
       {\rm for}\quad {^{208}}{\rm Bi},
\end{eqnarray}
respectively.  One may use Eq.(\ref{eq:107}) for the estimate of the
lower bound of the chameleon--matter coupling constant $\beta$. Since
the experiments on the measuring of the electron--neutron scattering
length have been carried out for liquid lead and bismuth
\cite{Kopecky1997} with densities $\rho_{\rm Pb} = 10.678\,{\rm
  g/cm^3}$ and $\rho_{\rm Bi} = 10.022\,{\rm g/cm^3}$, respectively,
in Fig.\,{\ref{fig:beta}} we plot the lower bound of the
chameleon--matter coupling $\beta$ at which the contributions of the
chameleon field are essential.

The minimal lower bound $\beta \ge 10^{19}$ is ten orders of magnitude
larger compared to the value $\beta < 5.8\times 10^8$, measured
recently in the qBounce experiments \cite{Jenke2014}. As a result, the
contribution of the chameleon field to the electron--scattering length
in the environment of the liquid lead and bismuth is negligible. Thus,
in order to obtain a tangible contribution of the chameleon to the
electron--neutron scattering length $b_{ne}$ or the squared charge
radius of the neutron $r^2_n$, the experiments should be carried out
in the environments with densities of order $\rho \sim 10^{-7}\,{\rm
  g/cm^3}$ or even smaller.

\subsection{Contributions of the chameleon field to the squared 
charge radius of the proton $r^2_p$ }

The results obtained above for the squared charge radius of the
neutron can be applied to the analysis of the contribution of the
chameleon (torsion) to the squared charge radius of the proton. Since
the interaction of the chameleon field with the proton is described by
the Lagrangian Eq.(\ref{eq:96}) with the replacement $n(x) \to p(x)$,
where $p(x)$ is the operator of the proton field, the contribution of
the chameleon field to the squared charge radius of the proton $r^2_p$
is defined by Eqs.(\ref{eq:104}) and (\ref{eq:45}) with the
replacement $r^2_n \to \delta r^2_p$, $m_n \to m_p$ and $m_e \to
m_{\mu}$, where $m_{\mu}$ is the mass of the $\mu^-$--meson
\cite{PDG2014}.

\begin{figure}
\centering \includegraphics[width=0.45\linewidth]{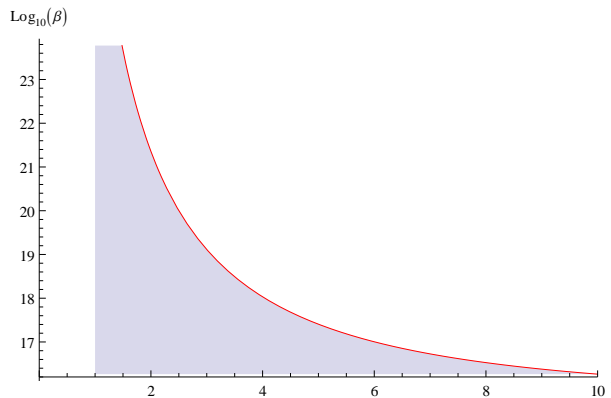}
\caption{The chameleon--matter coupling constant $\beta$ as a function
  of the index $n$, fitting the experimental value $\delta E_{2s - 2p}
  = \Delta E_{2s - 2p} - \Delta E_{2s - 2p}|_{r_p = 0.8768} =
  0.311\,{\rm meV}$, where $\Delta E_{2s - 2p} = 206.2949\,{\rm meV}$
  \cite{Pohl2010}, of the Lamb shift of muonic hydrogen
  \cite{Pohl2014}.The shaded area is excluded.}
 \label{fig:proton}
\end{figure}

The contribution of the chameleon field to the charge radius of the
proton has been recently investigated by Brax and Burrage
\cite{Brax2014}. According to Brax and Burrage \cite{Brax2014}, the
contribution of the chameleon field may solve the so--called ``the
proton radius anomaly''
\cite{Pohl2010,Batell2011,Pohl2013,Pohl2014}. As has been shown in
\cite{Pachucki1996}--\cite{Matynenko2008} the Lamb shift $\Delta E_{2s
  \to 2p}$ of the muonic hydrogen, calculated in QED with the account
for the nuclear effects, can be expressed in terms of the charge
radius of the proton $r_p$
\begin{eqnarray}\label{eq:108}
\Delta E_{2s \to 2p} = 209.9779(49) - 5.2262\,r^2_p + 0.0347\,r^3_p,
\end{eqnarray}
where $\Delta E_{2s \to 2p}$ and $r_p$ are measured in ${\rm meV}$ and
${\rm fm}$, respectively. The charge radius of the proton $r_p =
0.8768(69)\,{\rm fm}$, measured from the electronic hydrogen
\cite{Mohr2008} and agreeing well with the charge radius of the proton
$r_p = 0.879(8)\,{\rm fm}$, extracted from in the electron scattering
experiments \cite{Bernauer2010}. In turn, the experimental value of
the charge proton radius $r_p = 0.84087(39)\,{\rm fm}$, extracted from
the measurements of the Lamb shift of muonic hydrogen
\cite{Antognini2013}, is of about $4\,\%$ smaller compared to the
charge radius of the proton, measured from the electronic hydrogen
\cite{Batell2011}. The correction to the Lamb shift, caused by the
correction to the squared charge radius $\delta r^2_p$, is equal to
\begin{eqnarray}\label{eq:109}
\delta E_{2s \to 2p} = (- 5.2262 + 0.0521\,r_p)\,\delta r^2_p =
-5.180\,\delta r^2_p,
\end{eqnarray}
where we have set $r_p = 0.8768\,{\rm fm}$. The correction to the
charge radius of the proton, caused by low energy $\mu\,p$ scattering,
is equal to (see Eq.(\ref{eq:104}))
\begin{eqnarray}\label{eq:110}
\delta r^2_p = -
\frac{9}{4\pi^2}\,\frac{\beta^2}{m^2_{\phi}}\,\frac{m_{\mu}
  m_p}{M^2_{\rm Pl}}{\ell n}\Big(\frac{M_{\rm Pl}}{m_{\mu}}\Big) = -
6.617\times 10^{-15}\,\frac{\beta^2}{m^2_{\phi}},
\end{eqnarray}
where $m_{\mu} = 105.6584\,{\rm MeV}$ and $m_p = 938.2720\,{\rm MeV}$
are the muon and proton masses, respectively \cite{PDG2014}, and
$m_{\phi}$ and $\delta r^2_p$ are measured in ${\rm meV}$ and ${\rm
  fm^2}$, respectively. Substituting Eq.(\ref{eq:110}) into
Eq.(\ref{eq:109}) we express the correction to the Lamb shift of the
muonic hydrogen in terms of the parameters of the chameleon field
theory and the matter density $\rho_m$. We get
\begin{eqnarray}\label{eq:111}
\delta E_{2s \to 2p} = 3.428\times
10^{-14}\,\frac{\beta^2}{m^2_{\phi}} = 5.951\times
10^{-15}\,\frac{\beta^{\textstyle \frac{n}{n + 1}}}{n((n +
  1)}\,\Big(\frac{n M_{\rm Pl}\Lambda^3}{\rho_m}\Big)^{\textstyle
  \frac{n + 2}{n + 1}}
\end{eqnarray}
with $\delta E_{2s \to 2p} = 0.311\,{\rm meV}$ \cite{Pohl2014}.  In
Fig.\,{\ref{fig:proton}} we plot the coupling constant $\beta$ as a
function of the Ratra--Peebles index $n$ at the environment density
$\rho_m \approx 10\,{\rm g/cm^3}$ \cite{Chameleon1}. One may see that
for $\delta E_{2s \to 2p} = 0.311\,{\rm meV}$ the lower bound of the
chameleon--matter coupling constant $\beta$, at which the contribution
of the chameleon is tangible, is $\beta \ge 10^{16}$. This is seven
orders of magnitude larger compared to the recent experimental upper
bound $\beta < 5.8\times 10^8$ \cite{Jenke2014}.

\subsection{Contribution of the chameleon field to the neutron 
$\beta^-$--decay}

In this section we investigate the neutron $\beta^-$--decay, caused by
the interaction with the chameleon. This means that we investigate two
reactions: i) the neutron $\beta^-$--decay with an emission of the
chameleon $n \to p + e^- + \bar{\nu}_e + \phi$ and ii) the chameleon
induced neutron $\beta^-$--decay $\phi + n \to p + e^- +
\bar{\nu}_e$. Since formally these two reactions are related by
$k_{\phi} \to - k_{\phi}$, where $k_{\phi}$ is a 4--momentum of the
chameleon, we give the calculation of the amplitude of the neutron
$\beta^-$--decay $n \to p + e^- + \bar{\nu}_e + \phi$ with an emission
of the chameleon.

\subsubsection*{\bf Neutron $\beta^-$--decay with the chameleon particle 
in the final state $n \to p + e^- + \bar{\nu}_e + \phi$}

The calculation of the amplitude of such a decay we use the following
effective interactions
\begin{eqnarray}\label{eq:112}
{\cal L}_{\rm eff } &=& - \beta\,\frac{m_n}{M_{\rm
    Pl}}\,\bar{n}(x)n(x)\,\phi(x) - \beta\,\frac{m_p}{M_{\rm
    Pl}}\,\bar{p}(x)p(x)\,\phi(x)\nonumber\\ &&-
\frac{G_F}{\sqrt{2}}\,V_{ud}\,\Big\{[\bar{p}(x)\gamma_{\mu}(1 +
  \lambda \gamma^5)n(x)] + \frac{\kappa}{2 M}
\partial^{\nu}[\bar{p}(x)\sigma_{\mu\nu}n(x)]\Big\}[\bar{e}^-(x)\gamma^{\mu}(1
  - \gamma^5)\nu_e(x)],
\end{eqnarray}
where $G_F = 1.1664\times 10^{-1}\,{\rm MeV^{-2}}$ is the Fermi
coupling constant, $V_{ud} = 0.97427(15)$ is the
Cabibbo--Kobayashi--Maskawa (CKM) quark mixing matrix element
\cite{PDG2014}, $\lambda = - 1.2750(9)$ is the axial coupling constant
\cite{Abele2008} (see also\cite{Ivanov2013a}) and $\kappa = \kappa_p -
\kappa_n = 3.7058$ is the isovector anomalous magnetic moment of the
nucleon, defined by the anomalous magnetic moments of the proton
$\kappa_p = 1.7928$ and the neutron $\kappa_n = - 1.9130$ and measured
in nuclear magneton \cite{PDG2014}, $M = (m_n + m_p)/2$ is the nucleon
average mass.

\begin{figure}
 \includegraphics[width=0.55\linewidth]{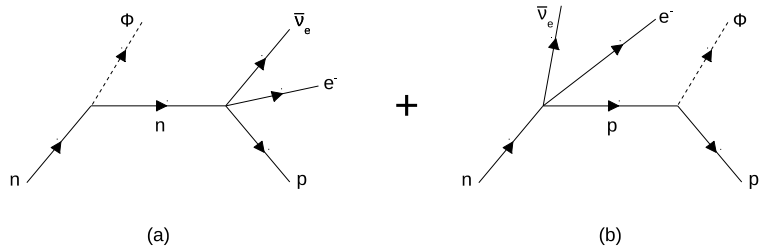}
\caption{Feynman diagrams of the neutron $\beta^-$--decay with an
  emission of the chameleon $n \to p + e^- + \bar{\nu}_e + \phi$.}
 \label{fig:feynman}
\end{figure}

The Feynman diagrams of the amplitude of the neutron $\beta^-$--decay
$n \to p + e^- + \bar{\nu}_e + \phi$ are shown in
Fig.\,\ref{fig:feynman}.  For the calculation of the analytical
expression of the decay amplitude we follow \cite{Ivanov2013a} and
carry out it in the rest frame of the neutron, keeping the
contributions of the terms to order $1/M$. The energy spectrum and
angular distribution of the neutron $\beta^-$--decay with polarised
neutron and unpolarised proton and electron we may write in the
following general form
\begin{eqnarray}\label{eq:113}
\hspace{-0.3in}\frac{d^{12}\lambda_{n\phi}}{d\Gamma^{12}} =
\frac{G^2_F|V_{ud}|^2}{4 m_n}\beta^2\frac{M^2}{M^2_{\rm
    Pl}}\overline{|M(n\to p\,e^-\,\bar{\nu}_e\,\phi)|^2}\,F(E_e, Z =
1)\,\Phi(\vec{k}_e,\vec{k}_{\nu},\vec{k}_{\phi})\,(2\pi)^4\,\delta^{(4)}(k_n
- k_p - k_e - k_{\nu} - k_{\phi}),
\end{eqnarray}
where $F(E_e, Z = 1)$ is the relativistic Fermi function, describing
the final--state electron--proton Coulomb interaction
\cite{Ivanov2013}, $d^{12}\Gamma$ is the phase--volume of the decay
final state
\begin{eqnarray}\label{eq:114}
d\Gamma^{12} = \frac{d^3k_p}{(2\pi)^3 2 E_p}\frac{d^3k_e}{(2\pi)^3 2
  E_e}\frac{d^3k_{\nu}}{(2\pi)^3 2 E_{\nu}}\frac{d^3k_{\phi}}{(2\pi)^3
  2 E_{\phi}}
\end{eqnarray}
and $k_j$ for $j = n, p, e, \nu$ and $\phi$ is a 4--momentum of the
neutron and the decay particles, respectively. The factor
$\Phi(\vec{k}_e,\vec{k}_{\nu},\vec{k}_{\phi})$ is the contribution of
the phase--volume, taking into account the terms of order
$1/M$. Following \cite{Ivanov2013a} one obtains
\begin{eqnarray}\label{eq:115}
\Phi(\vec{k}_e,\vec{k}_{\nu},\vec{k}_{\phi}) = 1 + \frac{3}{M}\Big(E_e
- \frac{\vec{k}_e\cdot \vec{k}_{\nu}}{E_{\nu}}\Big) +
\frac{3}{M}\Big(E_{\phi} - \frac{
  \vec{k}_{\phi}\cdot \vec{k}_{\nu}}{E_{\nu}}\Big) +
\frac{2}{M}\,\frac{E_e E_{\phi} - \vec{k}_e\cdot
  \vec{k}_{\phi}}{E_{\nu}}.
\end{eqnarray}
The last two terms define the deviation from the phase--volume factor,
calculated in \cite{Ivanov2013a} for the neutron $\beta^-$--decay $n
\to p + e^- + \bar{\nu}_e$. Taking into account the phase--volume
factor $\Phi(\vec{k}_e,\vec{k}_{\nu},\vec{k}_{\phi})$ we may carry out
the integration over the phase--volume of the $n \to p + e^- +
\bar{\nu}_e + \phi$ decay, neglecting the contribution of the kinetic
energy of the proton.

Then, $\overline{|M(n\to p\,e^-\,\bar{\nu}_e\,\phi)|^2}$ is the
squared absolute value of the decay amplitude, summed over the
polarisation of the decay electron and proton.
The analytical expression of the amplitude $M(n\to
p\,e^-\,\bar{\nu}_e\,\phi)$ is defined by
\begin{eqnarray}\label{eq:116}
\hspace{-0.3in}M(n\to p\,e^-\,\bar{\nu}_e\,\phi) &=&
\Big[\bar{u}_p(k_p,\sigma_p)\,\frac{1}{m_p - \hat{k}_p + \hat{k}_{\phi} -
  i0}\,O^{(-)}_{\mu}\, u_n(k_n,\sigma_n)\Big]\Big[\bar{u}_e(k_e,\sigma_e)\gamma^{\mu}(1
  - \gamma^5)v_{\bar{\nu}_e}(k_{\nu},+ \frac{1}{2})\Big]\nonumber\\
\hspace{-0.3in} &+&
\Big[\bar{u}_p(k_p,\sigma_p)\,O^{(+)}_{\mu}\,\frac{1}{m_n -
  \hat{k}_n + \hat{k}_{\phi} -
  i0}u_n(k_n,\sigma_n)\Big]\Big[\bar{u}_e(k_e,\sigma_e)\gamma^{\mu}(1 -
  \gamma^5)v_{\bar{\nu}_e}(k_{\nu},+ \frac{1}{2})\Big],
\end{eqnarray}
where $u_j(k_j,\sigma_j)$ for $j = n, p$ and $e$ are the Dirac
bispinors of fermions with polarisations $\sigma_j$,
$v_{\bar{\nu}_e}(k_{\nu}, +1/2)$ is the Dirac bispinor of the electron
antineutrino and $O_{\mu}$ is defined by \cite{Ivanov2013a}
\begin{eqnarray}\label{eq:117} 
O^{(\pm)}_{\mu}(k_p,k_n) = \gamma_{\mu}(1 + \lambda \gamma^5) +
i\,\frac{\kappa}{2 M}\,\sigma_{\mu\nu}(k_p - k_n \pm k_{\phi})^{\nu}.
\end{eqnarray}
In the accepted approximation the amplitude Eq.(\ref{eq:115}) can be
defined by the expression
\begin{eqnarray}\label{eq:118}
\hspace{-0.3in}M(n\to p\,e^-\,\bar{\nu}_e\,\phi) &=&
\frac{2}{E_{\phi}}\, [\bar{u}_p(k_p,\sigma_p)\,{\cal O}_{\mu}
  u_n(k_n,\sigma_n)]\Big[\bar{u}_e(k_e,\sigma_e)\gamma^{\mu}(1 -
  \gamma^5)v_{\bar{\nu}_e}(k_{\nu},+ \frac{1}{2})\Big]
=\nonumber\\ &=&\frac{2}{E_{\phi}}[\bar{u}_p(k_p,\sigma_p)\,{\cal O}_0
  u_n(k_n,\sigma_n)]\Big[\bar{u}_e(k_e,\sigma_e)\gamma^0(1 -
  \gamma^5)v_{\bar{\nu}_e}(k_{\nu},+ \frac{1}{2})\Big]\nonumber\\ &-&
\frac{2}{E_{\phi}}[\bar{u}_p(k_p,\sigma_p)\,\vec{{\cal O}}\,
  u_n(k_n,\sigma_n)]\cdot \Big[\bar{u}_e(k_e,\sigma_e)\vec{\gamma}\,(1
  - \gamma^5)v_{\bar{\nu}_e}(k_{\nu},+ \frac{1}{2})\Big]
\end{eqnarray}
where ${\cal O}_{\mu}$ is given by
\begin{eqnarray}\label{eq:119} 
{\cal O}_{\mu} = \gamma_{\mu}(1 + \lambda \gamma^5) +
i\,\frac{\kappa}{2 M}\,\sigma_{\mu\nu}(k_p - k_n )^{\nu} -
\frac{k_{\phi\mu}}{2M} - i\,\frac{\lambda}{2 M}\,\sigma_{\mu\nu}
\gamma^5 k^{\nu}_{\phi}
\end{eqnarray}
and  the matrices ${\cal O}_0$ and $\vec{{\cal O}}$, taken to order
$1/M$, are equal to
\begin{eqnarray}\label{eq:120}
 \hspace{-0.3in} O^0 =\left(\begin{array}{ccc} 1 - {\displaystyle
     \frac{E_{\phi}}{2M} + \frac{\lambda}{2 M}\,(\vec{\sigma}\cdot
     \vec{k}_{\phi})}& {\displaystyle \lambda + \frac{\kappa}{2
       M}\,(\vec{\sigma}\cdot \vec{k}_p)}\\ {\displaystyle - \lambda +
     \frac{\kappa}{2 M}\,(\vec{\sigma}\cdot \vec{k}_p) } & - 1 -
   {\displaystyle \frac{E_{\phi}}{2M} + \frac{\lambda}{2
       M}\,(\vec{\sigma}\cdot \vec{k}_{\phi})}\\
    \end{array}\right)
\end{eqnarray}
and 
\begin{eqnarray}\label{eq:121}
 \hspace{-0.3in}&& \vec{O} =\left(\begin{array}{ccc} {\displaystyle
     \lambda \vec{\sigma}\Big(1 - \frac{E_{\phi}}{2M}\Big) -
     \frac{\vec{k}_{\phi}}{2M} + i\,\frac{\kappa}{2
       M}\,(\vec{\sigma}\times \vec{k}_p)} & {\displaystyle
     \vec{\sigma}\,\Big(1 - \frac{\kappa}{2 M}\,E_0\Big) -
     i\,\frac{\lambda}{2 M}\,(\vec{\sigma}\times
     \vec{k}_{\phi})}\\ {\displaystyle -\,\vec{\sigma}\,\Big(1 +
     \frac{\kappa}{2 M}\,E_0\Big) - i\,\frac{\lambda}{2
       M}\,(\vec{\sigma}\times \vec{k}_{\phi})} & {\displaystyle -
     \lambda \vec{\sigma}\Big(1 + \frac{E_{\phi}}{2M}\Big) -
     \frac{\vec{k}_{\phi}}{2M} + i \frac{\kappa}{2
       M}\,(\vec{\sigma}\times \vec{k}_p)} \\
    \end{array}\right),\nonumber\\
 \hspace{-0.3in}&&
\end{eqnarray}
where $E_0 = ((m_n - m_{\phi})^2 - m^2_p + m^2_e)/2 (m_n - m_{\phi})$
is the end--point energy of the electron--energy spectrum. The
matrices $O^0$ and $\vec{O}$ are defined to order $1/M$ only
\cite{Ivanov2013a} . For the calculation of the amplitude of the
$\beta^-$--decay of the neutron we use the Dirac bispinorial wave
functions of the neutron and the proton
\begin{eqnarray}\label{eq:122}
 \hspace{-0.3in} u_n(\vec{0},\sigma_n) = \sqrt{2
   m_n}\Big(\begin{array}{c}\varphi_n \\ 0
 \end{array}\Big) \quad,\quad u_p(\vec{k}_p,\sigma_p) = \sqrt{E_p +
 m_p}\left(\begin{array}{c}\varphi_p \\ {\displaystyle
     \frac{\vec{\sigma}\cdot \vec{k}_p}{E_p + m_p}\,\varphi_p}
 \end{array}\right),
\end{eqnarray}
where $\varphi_n$ and $\varphi_p$ are the Pauli spinor functions of
the neutron and proton, respectively.  For the energy spectrum and
angular distribution of the neutron $\beta^-$--decay with polarised
neutron and unpolarised proton and electron we may write in the
following general form
\begin{eqnarray}\label{eq:123}
\hspace{-0.3in}&&\frac{d^{12}\lambda_{n\phi}}{d\Gamma^{12}} =
32\,m_p\,E_e
E_{\nu}\,\frac{G^2_F|V_{ud}|^2}{E^2_{\phi}}\beta^2\frac{M^2}{M^2_{\rm
    Pl}}\,F(E_e, Z = 1)\,(2\pi)^4\,\delta^{(4)}(k_n - k_p - k_e -
k_{\nu} - k_{\phi})\,(1 + 3 \lambda^2)\,\zeta(E_e)\nonumber\\
\hspace{-0.3in}&&\times\,\Big\{1 + a(E_e)\,\frac{\vec{k}_e\cdot
  \vec{k}_{\nu}}{E_e E_{\nu}} + A(E_e)\,\frac{\vec{\xi}_n\cdot \vec{k}_e}{E_e} +
B(E_e)\, \frac{\vec{\xi}_n\cdot \vec{k}_{\nu}}{E_{\nu}} +
K_n(E_e)\,\frac{(\vec{\xi}_n\cdot \vec{k}_e)(\vec{k}_e\cdot
  \vec{k}_{\nu})}{E^2_e E_{\nu}}+ Q_n(E_e)\,\frac{(\vec{\xi}_n\cdot
  \vec{k}_{\nu})(\vec{k}_e\cdot \vec{k}_{\nu})}{ E_e E^2_{\nu}} \nonumber\\
\hspace{-0.3in}&&+ D(E_e)\,\frac{\vec{\xi}_n\cdot (\vec{k}_e\times
  \vec{k}_{\nu})}{E_e E} - 3\,\frac{1 - \lambda^2}{1 +
  3\lambda^2}\,\frac{E_e}{M}\,\Big(\frac{(\vec{k}_e\cdot
  \vec{k}_{\nu})^2}{E^2_e E^2_{\nu}} -
\frac{1}{3}\,\frac{k^2_e}{E^2_e}\,\Big) + \frac{1}{M}\,\frac{1}{1 +
  3\lambda^2}\,F_{\phi}\Big\},
\end{eqnarray}
where $k_e = \sqrt{E^2_e - m^2_e}$ is the absolute value of the
electron 3--momentum and $\vec{\xi}_n$ is the unit polarisation vector
of the neutron $|\vec{\xi}_n| = 1$. The correlation coefficients
$\zeta(E_e)$, $a(E_e)$, $A(E_e)$, $B(E_e)$, $K_n(E_e)$, $Q_n(E_e)$ and
$D(E_e)$ can taken from \cite{Ivanov2013a} at the neglect of the
radiative corrections. The correction coefficient  $F_{\phi}$ we may
represent in the following form
\begin{eqnarray}\label{eq:124}
F_{\phi} = F^{(1)}_{\phi} +  F^{(2)}_{\phi} +  F^{(3)}_{\phi},
\end{eqnarray}
where the correlation coefficients $F^{(j)}_{\phi}$ for $j = 1,2,3$
are defined by i) the contributions of the terms, depending on the
energy and 3--momentum of the chameleon particle in the matrices
${\cal O}_0$ and $\vec{\cal O}$ given by Eqs.(\ref{eq:119}) and
(\ref{eq:120}), ii) the dependence of the 3--momentum of the proton
$\vec{k}_p$ on the 3--momentum of the chameleon particle, caused by
the 3--momentum conservation $\vec{k}_p = - \vec{k}_e - \vec{k}_{\nu}
- \vec{k}_{\phi}$ (see Eqs.(A.16) and (A.17) in Appendix A of
Ref.\cite{Ivanov2013a}) and iii) the contributions of the
phase--volume factor Eq.(\ref{eq:115}), respectively. The analytical
expressions of these correlation coefficients are equal to
\begin{eqnarray}\label{eq.125}
\hspace{-0.3in}F^{(1)}_{\phi} &=& (1 + 3 \lambda^2)\,E_{\phi} - (1 -
  \lambda^2)\,E_{\phi}\,\frac{\vec{k}_e\cdot \vec{k}_{\nu}}{E_e
    E_{\nu}} + 2\,\lambda^2\,E_{\phi}\,\frac{\vec{\xi}_n\cdot
    \vec{k}_e}{E_e} - 2\,\lambda^2\,E_{\phi}\,\frac{\vec{\xi}_n\cdot
    \vec{k}_{\nu}}{E_{\nu}}\nonumber\\
 \hspace{-0.3in}&+& 2\,\lambda\,\frac{(\vec{\xi}_n\cdot
    \vec{k}_{\phi})(\vec{k}_e\cdot \vec{k}_{\nu})}{E_e
    E_{\nu}} - \lambda\,\frac{(\vec{\xi}_n\cdot
    \vec{k}_e)(\vec{k}_{\nu}\cdot \vec{k}_{\phi})}{E_e E_{\nu}} -
  \lambda\,\frac{(\vec{\xi}_n\cdot \vec{k}_{\nu})(\vec{k}_e\cdot
    \vec{k}_{\phi})}{E_e E_{\nu}},
\end{eqnarray}
\begin{eqnarray}\label{eq:126}
\hspace{-0.3in}F^{(2)}_{\phi} &=& (\lambda^2 - 2(\kappa +
1)\,\lambda)\,\frac{\vec{k}_e\cdot \vec{k}_{\phi}}{E_e} + (\lambda^2 +
2(\kappa + 1)\,\lambda)\,\frac{\vec{k}_{\nu}\cdot
  \vec{k}_{\phi}}{E_{\nu}} + (2\kappa + 1)\,\lambda\,(\vec{\xi}_n\cdot
\vec{k}_{\phi}) - \lambda\,\frac{(\vec{\xi}_n\cdot
  \vec{k}_{\phi})(\vec{k}_e\cdot \vec{k}_{\nu})}{E_e
  E_{\nu}}\nonumber\\ \hspace{-0.3in}&-& (\lambda^2 + (\kappa +
1)\,\lambda + (\kappa + 1))\,\frac{(\vec{\xi}_n\cdot
  \vec{k}_e)(\vec{k}_{\nu}\cdot \vec{k}_{\phi})}{E_e E_{\nu}} +
(\lambda^2 - (\kappa + 1)\lambda + (\kappa +
1))\,\frac{(\vec{\xi}_n\cdot \vec{k}_{\nu})(\vec{k}_e\cdot
  \vec{k}_{\phi})}{E_e E_{\nu}}
\end{eqnarray}
and
\begin{eqnarray}\label{eq:127}
F^{(3)}_{\phi} = \Big[3\,\Big(E_{\phi} - \frac{ \vec{k}_{\phi}\cdot
      \vec{k}_{\nu}}{E_{\nu}}\Big) + 2\,\frac{E_e E_{\phi} -
      \vec{k}_e\cdot \vec{k}_{\phi}}{E_{\nu}}\Big]\Big[(1 -
    \lambda^2)\,\frac{\vec{k}_e\cdot \vec{k}_{\nu}}{E_e E_{\nu}} - 2\lambda(1 + \lambda)\,\frac{\vec{\xi}_n\cdot \vec{k}_e}{E_e} - 2\lambda(1 - \lambda)\,\frac{\vec{\xi}_n\cdot \vec{k}_{\nu}}{E_{\nu}}\Big].
\end{eqnarray}
Now we may integrate over the phase--volume of the $n \to p + e^- +
\bar{\nu}_e + \phi$ decay. First of all we integrate over the
3--momentum of the proton $\vec{k}_p$. As has been mention above, we
make such an integration at the neglect of the kinetic energy of the
proton. Then, since one can hardly observe the dependence of the
energy spectrum and the angular distribution on the direction of the
3--momentum of the chameleon, we make the integration over the
3--momentum of the chameleon $\vec{k}_{\phi}$. The obtained energy
spectrum and angular distribution is
\begin{eqnarray}\label{eq:128}
\hspace{-0.3in}&&\frac{d^7\lambda_{n\phi}(E_e,E_{\nu},E_{\phi},\vec{k}_e,\vec{k}_{\nu},
  \vec{\xi}_n)}{dE_e dE_{\nu} dE_{\phi}d\Omega_e d\Omega_{\nu}} =
\frac{G^2_F|V_{ud}|^2}{32\pi^7 E_{\phi}}\beta^2\frac{M^2}{M^2_{\rm
    Pl}}\,F(E_e, Z = 1)\,\delta(E_0 - E_e - E_{\nu} -
E_{\phi})\,k_e E_e\,E^2_{\nu} \,(1 + 3
\lambda^2)\,\zeta_{\phi}(E_e)\nonumber\\
\hspace{-0.3in}&&\times\,\Big\{1 + a_{\phi}(E_e)\,\frac{\vec{k}_e\cdot
  \vec{k}_{\nu}}{E_e E_{\nu}} + A_{\phi}(E_e)\,\frac{\vec{\xi}_n\cdot
  \vec{k}_e}{E_e} + B_{\phi}(E_e)\, \frac{\vec{\xi}_n\cdot
  \vec{k}_{\nu}}{E_{\nu}} + K_n(E_e)\,\frac{(\vec{\xi}_n\cdot
  \vec{k}_e)(\vec{k}_e\cdot \vec{k}_{\nu})}{E^2_e E_{\nu}}+
Q_n(E_e)\,\frac{(\vec{\xi}_n\cdot \vec{k}_{\nu})(\vec{k}_e\cdot
  \vec{k}_{\nu})}{ E_e E^2_{\nu}} \nonumber\\
\hspace{-0.3in}&&+ D(E_e)\,\frac{\vec{\xi}_n\cdot (\vec{k}_e\times
  \vec{k}_{\nu})}{E_e E_{\nu}} - 3\,\frac{1 - \lambda^2}{1 +
  3\lambda^2}\,\frac{E_e}{M}\,\Big(\frac{(\vec{k}_e\cdot
  \vec{k}_{\nu})^2}{E^2_e E^2_{\nu}} -
\frac{1}{3}\,\frac{k^2_e}{E^2_e}\,\Big)\Big\}.
\end{eqnarray}
The correlation coefficients $\zeta_{\phi}(E_e)$, $a_{\phi}(E_e)$,
$A_{\phi}(E_e)$ and $B_{\phi}(E_e)$ are equal to
\begin{eqnarray}\label{eq:129}
\hspace{-0.3in}\zeta_{\phi}(E_e) &=& \bar{\zeta}(E_e) +
\frac{E_{\phi}}{M},\nonumber\\
\hspace{-0.3in}a_{\phi}(E_e) &=& \bar{a}(E_e) + a_0\,\Big(1 +
2\,\frac{E_e}{E_{\nu}}\Big)\,\frac{E_{\phi}}{M}\nonumber\\
\hspace{-0.3in}A_{\phi}(E_e) &=& \bar{A}(E_e) + \Big( - 2\lambda +
A_0\,\Big(1 +
2\,\frac{E_e}{E_{\nu}}\Big)\Big)\,\frac{E_{\phi}}{M},\nonumber\\
\hspace{-0.3in}B_{\phi}(E_e) &=& \bar{B}(E_e) + \Big( - 2\lambda +
B_0\,\Big(1 + 2\,\frac{E_e}{E_{\nu}}\Big)\Big)\, \frac{E_{\phi}}{M},
\end{eqnarray}
where $a_0 = (1 - \lambda^2)/(1 + 3\lambda^2)$, $A_0 = - 2\lambda(1 +
\lambda)/(1 + 3\lambda^2)$ and $B_0 = - 2\lambda(1 - \lambda)/(1 +
3\lambda^2)$ \cite{Abele2008}(see also \cite{Ivanov2013a}). The
correlation coefficients $\bar{\zeta}(E_e)$, $\bar{a}(E_e)$,
$\bar{A}(E_e)$ and $\bar{B}(E_e)$ are calculated in
\cite{Ivanov2013a} by taking into account the contributions of the
weak magnetism and the proton recoil to order $1/M$ but without
radiative corrections.

\begin{figure}
 \includegraphics[width=0.45\linewidth]{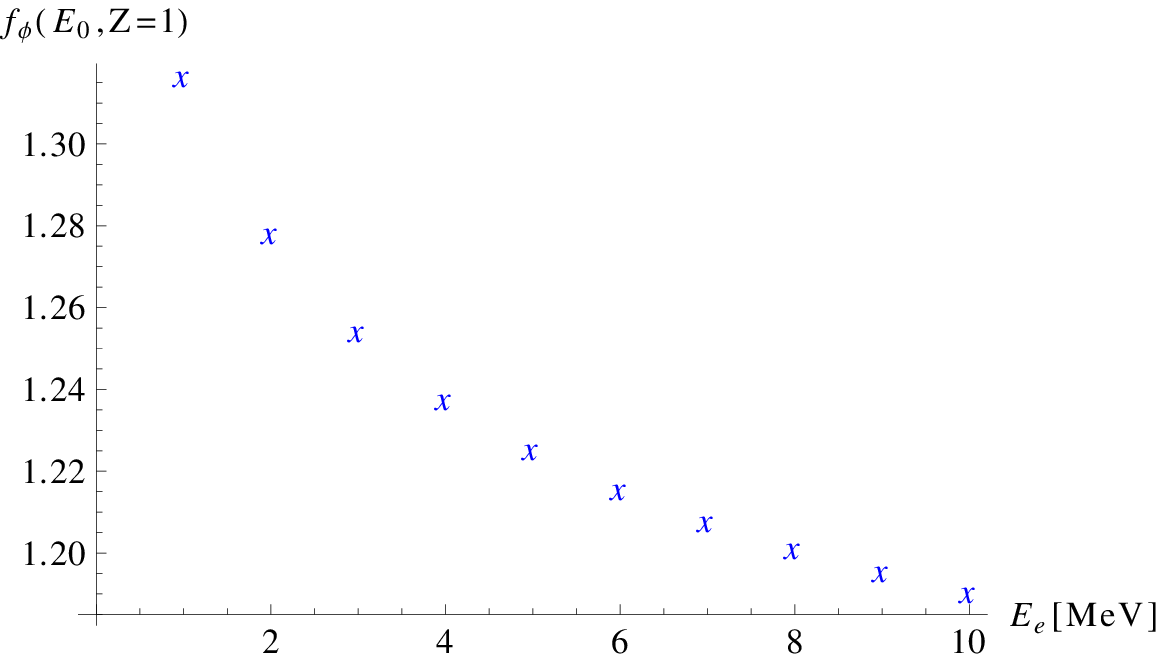}
\includegraphics[width=0.45\linewidth]{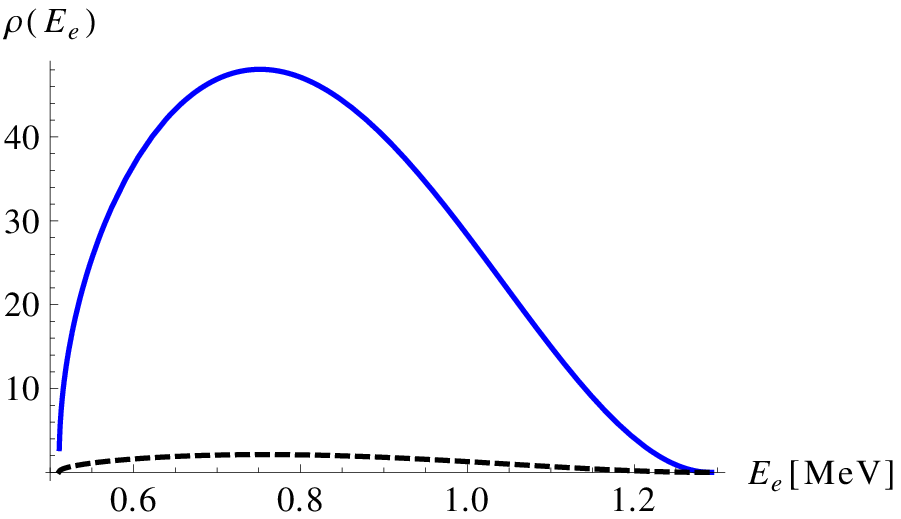}
\caption{(left) The values of the Fermi integral $f_{\phi}(E_0, Z =
  1)$ as a function of the index $n$ for $n = 1,2,\ldots, 10$. (right)
  The electron--energy spectra $\rho_{\phi}(E_e)$ (continuous lines)
  and $\rho_{\beta^-_c}(E_e)$ (dashed line) of the neutron
  $\beta^-$--decay with and without an emission of a chameleon,
  respectively. The densities $\rho_{\phi}(E_e)$ depend slightly on
  the index $n$ and are represented by only one continuous blue line.}
 \label{fig:rho_phi_spectrum}
\end{figure}

The rate of the decay $n \to p + e^- + \bar{\nu}_e + \phi$ diverges
logarithmically at $E_{\phi} \to 0$.  We regularise the
logarithmically divergent integral by the chameleon mass
$m_{\phi}$. As result we get
\begin{eqnarray}\label{eq:130}
\hspace{-0.3in}\lambda_{n\phi} =
\frac{G^2_F|V_{ud}|^2}{2\pi^3}\,\frac{\beta^2}{\pi^2}\,\frac{M^2}{M^2_{\rm
    Pl}}\,f_{\phi}(E_0, Z = 1),
\end{eqnarray}
where $f_{\phi}(E_0,Z = 1)$ is the Fermi integral
\begin{eqnarray}\label{eq:131}
\hspace{-0.3in}f_{\phi}(E_0, Z = 1) =
\int^{E_0}_{m_e}dE_e\,E_e\,\sqrt{E^2_e - m^2_e}\,(E_0 - E_e)^2\Big[{\ell
    n}\Big(\frac{E_0 - E_e}{m_{\phi}}\Big) - \frac{3}{2} +
  \frac{1}{3}\,\frac{E_0 - E_e}{M}\Big]\, \bar{\zeta}(E_e)\,F(E_e, Z =
1),
\end{eqnarray}
where $\bar{\zeta}(E_e)$ is given by (see Eq.(\ref{eq:7}) of
Ref.\cite{Ivanov2013a})
\begin{eqnarray}\label{eq:132}
\hspace{-0.3in}&&\bar{\zeta}(E_e) = 1 + \frac{1}{M}\,\frac{1}{1 + 3
  \lambda^2}\, \Big[- 2\,\lambda\Big(\lambda - (\kappa + 1)\Big)\,E_0
  + \Big(10 \lambda^2 - 4(\kappa + 1)\, \lambda + 2\Big)\,E_e - 2
  \lambda\,\Big(\lambda - (\kappa + 1)\Big)\,\frac{m^2_e}{E_e}\Big].
\end{eqnarray}
In Fig.\,\ref{fig:rho_phi_spectrum} we plot the electron--energy
spectrum $\rho_{\phi}(E_e)$ of the neutron $\beta^-$--decay with an
emission of the chameleon, defined by
\begin{eqnarray}\label{eq:133}
\hspace{-0.3in}\rho_{\phi}(E_e) = E_e\,\sqrt{E^2 - m^2_e}\,(E_0 -
E_e)^2\Big[{\ell n}\Big(\frac{E_0 - E_e}{m_{\phi}}\Big) - \frac{3}{2}
  + \frac{1}{3}\,\frac{E_0 - E_e}{M}\Big]\,
\bar{\zeta}(E_e)\,\frac{F(E_e, Z = 1)}{f(E_0, Z = 1)},
\end{eqnarray}
where $f(E_0, Z = 1)$ is the Fermi integral calculated in
\cite{Ivanov2013a}, and compare it with the electron--energy spectrum
$\rho_{\beta^-_c}(E_e)$ of the neutron $\beta^-$--decay calculated in
\cite{Ivanov2013a} (see Eq.(D-59) of Ref.\cite{Ivanov2013a}). The
chameleon mass $m_{\phi}$ is determined at the local density $\rho
\simeq 1.19\times 10^{-11}\,{\rm g/cm^3} = 5.12\times 10^{-17}\,{\rm
  MeV^4}$. This is the density of air at room temperature and pressure
$P \simeq 10^{-5}\,{\rm mbar} $
\cite{Serebrov2008,Pichlmaier2010}. For $\beta < 5.8\times 10^8$ we
obtain $\lambda_{n\phi} < 2\times 10^{-34}\,{\rm s^{-1}}$ and the
branching ratio ${\rm BR}_{n\phi} < 1.8 \times 10^{-31}$.

\subsubsection*{\bf Neutron $\beta^-$--decay $\phi + n \to p + e^- + \bar{\nu}_e$, induced by the chameleon field}

\begin{figure}
 \includegraphics[width=0.50\linewidth]{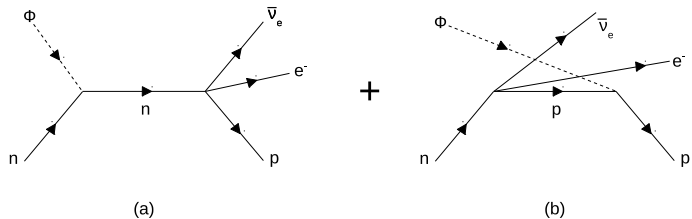}
\caption{Feynman diagrams for the reaction $\phi + n
    \to p + e^- + \bar{\nu}_e$}
 \label{fig:feynmanch}
\end{figure}

For the calculation of the amplitude $M(\phi\,n \to
p\,e^-\,\bar{\nu}_e)$ of the induced neutron $\beta^-$--decay $\phi +
n \to p + e^- + \bar{\nu}_e$ we may use the amplitude
Eq.(\ref{eq:116}) with the replacement $k_{\phi} \to - k_{\phi}$,
where $k_{\phi}$ is a 4--momentum of the chameleon. The Feynman
diagrams for the chameleon--induced neutron $\beta^-$--decay is shown
in Fig.\,\ref{fig:feynmanch}. 
The cross section for the induced neutron $\beta^-$--decay is
\begin{eqnarray}\label{eq:134}
\hspace{-0.3in}&&\sigma_{\phi\,n \to p\,e^-\,\bar{\nu}_e}(E_{\phi}) =
\frac{1}{4 m_n E_{\phi}}\int \overline{|M(\phi\,n \to
  p\,e^-\,\bar{\nu}_e)|^2}\,(2\pi)^4\,\delta^{(4)}(k_n + k_{\phi} - k_p -
k_e - k_{\nu})\,\frac{d^3k_p}{(2\pi)^3 2 E_p}\frac{d^3k_e}{(2\pi)^3 2
  E_e}\frac{d^3k_{\nu}}{(2\pi)^3 2 E_{\nu}},\nonumber\\
\hspace{-0.3in}&&
\end{eqnarray}
where the contribution of the electron--proton final--state Coulomb
interaction is not important and neglected.  Using the results,
obtained in previous subsection, we get
\begin{eqnarray}\label{eq:135}
\hspace{-0.3in}\sigma_{\phi\,n \to p\,e^-\,\bar{\nu}_e}(E_{\phi}) =
(1 + 3\lambda^2)\,\frac{G^2_F|V_{ud}|^2}{2\pi^3
  E^3_{\phi}}\,\beta^2\,\frac{M^2}{M^2_{\rm Pl}} \int^{E_0 +
  E_{\phi}}_{m_e}dE_e\,E_e\,\sqrt{E^2 - m^2_e}\,(E_0 + E_{\phi} -
E_e)^2.
\end{eqnarray}
Integrating over $E_e$ we obtain
\begin{eqnarray}\label{eq:136}
\hspace{-0.3in}\sigma_{\phi\,n \to p\,e^-\,\bar{\nu}_e}(E_{\phi}) &=&
(1 + 3\lambda^2)\,\frac{G^2_F|V_{ud}|^2}{2\pi^3
}\,\beta^2\,\frac{M^2}{M^2_{\rm Pl}} \frac{(E_0 + E_{\phi})^5}{30
  E^3_{\phi}}\Bigg\{\Bigg(1 - \frac{9}{2}\,\frac{m^2_e}{(E_0 +
  E_{\phi})^2} - 4\,\frac{m^4_e}{(E_0 + E_{\phi})^4}\Bigg)\nonumber\\
\hspace{-0.3in}&&\times \sqrt{1 - \frac{m^2_e}{(E_0 + E_{\phi})^2}} +
\frac{15}{2}\,\frac{m^4_e}{(E_0 + E_{\phi})^4}\,{\ell
  n}\Bigg(\frac{E_0 + E_{\phi}}{m_e} +\sqrt{\frac{(E_0 +
    E_{\phi})^2}{m^2_e} - 1}\;\Bigg)\Bigg\}.
\end{eqnarray}
Using the results, obtained in \cite{Brax2012} (see also
\cite{Baum2014}), we may analyse the quantity
\begin{eqnarray}\label{eq:137}
\lambda_{\phi n} = \int^{\infty}_0 dE_{\phi}\,\Phi_{\rm
  ch}(E_{\phi})\,\sigma_{\phi\,n \to p\,e^-\,\bar{\nu}_e}(E_{\phi}),
\end{eqnarray}
which defines the number of transitions $n \to p\,e^-\bar{\nu_e}$ per
second, induced by the chameleon, where $\Phi_{\rm ch}(E_{\phi})$ is
the number of solar chameleons per ${\rm eV\cdot s\cdot cm^2}$,
normalised to $10\,\%$ of the solar luminosity per unite area
$L_{\odot}/4\pi R_{\odot} = 3.9305\times 10^{22}\,{\rm
  eV\,s^{-1}\,cm^{-2}} = 0.01007\,{\rm eV^4}$ \cite{Brax2012}, where
$L_{\odot} = 2.3893\times 10^{45}\,{\rm eV\,s^{-1}}$ and $R_{\odot} =
6.9551(4)\times 10^{10}\,{\rm cm}$ are the total luminosity and the
radius of the Sun \cite{PDG2014}. Following \cite{Brax2012} we obtain
that $\lambda_{\phi n} < 10^{-41}\,{\rm s^{-1}}$ for $\beta <
5.8\times 10^8$ \cite{Jenke2014}.

\section{Conclusion}
\label{sec:conclusion}

We have developed the results, obtained in \cite{Ivanov2014}, where
there was shown that the chameleon field can serve also as a source
for a torsion field and low--energy torsion--neutron interactions,
where the torsion field is determined by a gradient of the chameleon
one.  Following Hojman {\it et al.} \cite{Hojman1978} we have extended
the Einstein gravitational theory with the chameleon field to a
version of the Einstein--Cartan gravitational one with a torsion
field. For the inclusion of the torsion field we have used a modified
form of local gauge invariance in the Weinberg--Salam electroweak
model with minimal coupling and derived the Lagrangians of the
electroweak and gravitational interactions with the chameleon
(torsion) field.

Gauge invariance of the torsion--photon interactions has been
explicitly checked by calculating the amplitudes of the two--photon
decay of the torsion (chameleon) field $\phi \to \gamma + \gamma$ and
the photon--torsion (chameleon) scattering $\gamma + \phi \to \phi +
\gamma$ or the Compton photon--torsion (chameleon) scattering. Unlike
the Compton--scattering, where photons scatter by free charged
particles with charged particles in the virtual intermediate states,
in the photon--torsion (chameleon) scattering a transition from an
initial $(\gamma\,\phi)$ state to a final $(\gamma\,\phi)$ goes
through the one--virtual photon exchange (see
Fig.\,\ref{fig:torsion2}a and Fig.\,\ref{fig:torsion2}b) and the local
${\cal L}_{\gamma\gamma\phi\phi}$ interaction (see
Fig.\,\ref{fig:torsion2}c). The Feynman diagram
Fig.\,\ref{fig:torsion2}d is self--gauge invariant due to the local
${\cal L}_{\gamma\gamma\phi}$ interaction. Gauge invariance has been
checked directly by a replacement of the one of the polarisation
vectors of the photons in the initial and final state by its
4-momentum. Since in these reactions the coupling constant of the
photon--torsion (chameleon) interaction is $g_{\rm eff} = \beta/M_{\rm
  Pl}$, in analogy with gauge invariance of photon--charge particles
interactions, where electric charge is a coupling constant -
unrenormalisable by any interactions, one may assert that the coupling
constant $g_{\rm eff} = \beta/M_{\rm Pl}$ should be also
unrenormalisable by any interactions. This may place some strict
constraints on possible mechanisms of the chameleon--matter coupling
constant $\beta/M_{\rm Pl}$ screening
\cite{Khoury2010,BraxCQG2013}. In this connection the Vainstein
mechanism, leading the screening of the coupling constant
$\beta/M_{\rm Pl}$ by the factor $1/\sqrt{Z}$, where $Z > 1$ is a
finite renormalisation constant of the wave function of the chameleon
field caused by a self--interaction of the chameleon field
\cite{BraxCQG2013} or some new higher derivative terms
\cite{Bloomfield2014}, is prohibited in such a version of the
Einstein--Cartan gravity with the chameleon field and torsion. Because
of the smallness of the constant $\beta^4/M^4_{\rm Pl} <
10^{-60}\,{\rm barn/eV^2}$, estimated for $\beta < 5.8\times 10^8$
\cite{Jenke2014}, the cross section for the photon--chameleon
scattering is extremely small and hardly may play any important
cosmological role at low energies, for example, for a formation of the
cosmological microwave background and so on. However, since in our
approach the coupling constant $\beta_{\gamma}/M_{\rm Pl}$ is fixed in
terms of the coupling constant $\beta/M_{\rm Pl}$, the recent
measurement of the upper bound $\beta < 5.8\times 10^8$ can make new
constraints on the photon--chameleon oscillations in the magnetic
field of the laboratory search for the chameleon field
\cite{Steffen2010,Schelpe2010,Rybka2010}.

In our approach the effective chameleon--photon coupling $g_{\rm eff}
= \beta/M_{\rm Pl}$ is equal to $g_{\rm eff} = \beta/M_{\rm Pl} <
2.4\times 10^{-10}\,{\rm GeV^{-1}}$, where we have used the
experimental upper bound of the chameleon--matter coupling constant
$\beta < 5.8\times 10^8$ \cite{Jenke2014}. The obtained upper bound
$g_{\rm eff} < 2.4\times 10^{-10}\,{\rm GeV^{-1}}$ is in qualitative
agreement with the upper bounds, estimated by Davis, Schelpe and Shaw
\cite{Davis2009}. Then, the constraints on $\beta$: $\beta < 1.9\times
10^7\;(n = 1)$, $\beta < 5.8\times 10^7\; (n = 2)$, $\beta < 2.0
\times 10^8\,(n = 3)$ and $\beta < 4.8 \times 10^8\,(n = 4)$, measured
recently by H. Lemmel {\it et al.} \cite{Lemmel2015} using the neutron
interferometer, may be used for more strict constraints on the
astrophysical sources of chameleons, investigated in
\cite{Brax2007}--\cite{Davis2009}.

Using the photon--torsion (chameleon) interaction we have estimated
the contributions of the chameleon field to the charged radii of the
neutron and proton. All tangible contributions can appear only for
$\beta \gg 10^8$. This, of course, is not compatible with recent
experimental data $\beta < 5.8\times 10^8$ by Jenke {\it et al.}
\cite{Jenke2014}. The branching ratio for the production of the
chameleon in the neutron $\beta^-$--decay $n \to p + e^- + \bar{\nu}_e
+ \phi$ is extremely small ${\rm Br}(n \to p\,e^-\,\bar{\nu}_e \,\phi)
< 1.8 \times 10^{-31}$. In turn, the half--life of the neutron
$T^{(\phi\,n)}_{1/2} ={\ell n}2/\lambda_{\phi n}$, caused by the
chameleon induced neutron $\beta^-$--decay $\phi + n \to p + e^- +
\bar{\nu}_e$, is extremely large $T^{(\phi\,n)}_{1/2} = {\ell
  n}2/\lambda_{\phi n} > 2 \times 10^{33}\,{\rm yr}$. Of course,
because of the neutron life--time $\tau_n = 880.3(1.1)\,{\rm s}$
\cite{PDG2014}, being in agreement with the recent theoretical value
$\tau_n = 879.6(1.1)\,{\rm s}$ \cite{Ivanov2013a}, the chameleon
induced neutron $\beta^-$--decay cannot be observed by a free
neutron. The experiment, which can give any meaningful result, can be
organised in a way, which is used for the detection of the
neutrinoless double $\beta$ decays \cite{Bahcall2004}. For example, it
is known that the isotope ${^{76}}{\rm Ge}$ is both stable with
respect to non--exotic weak, electromagnetic and nuclear decays and
are neutron--rich. It is unstable only with respect to the
neutrinoless double $\beta^-$--decay \cite{Bahcall2004}. The
experimental analysis of the low--bound on the half--life of
${^{76}}{\rm Ge}$ has been carried out by the GERDA Collaboration
Agostini {\it et al.}  \cite{Gerda2013a,Gerda2013b} by measuring the
energy spectrum of the electrons. The experimental lower bound has
been found to be equal to $T_{1/2} > 3\times 10^{25}\,{\rm yr}$
($90\,\%\,{\rm C.L.}$). We would like also to mention the experiments
on the proton decays, carried out by Super--Kamiokande
\cite{SuperKamiokande2009,SuperKamiokande2014}. For specific modes of
the proton decay, i.e. $p \to e^+ \pi^0$, $p \to \mu^+\pi^0$ and $p
\to \nu K^+$, for the half--life of the proton there have been found
the following lower bounds: $T_{1/2} > 5.7\times 10^{33}\,{\rm yr}$,
$T_{1/2} > 4.6\times 10^{33}\,{\rm yr}$ and $T_{1/2} > 4.1\times
10^{33}\,{\rm yr}$, respectively.

Since the lifetime of the neutron $\tau_n = 880.3(1.1)\,{\rm s}$
\cite{PDG2014}, one cannot analyse experimentally the
chameleon--induced $\beta^-$--decay on free neutrons. For the
experimental investigation of the chameleon--induced $\beta^-$--decay
one may propose the following reaction
\begin{eqnarray}\label{eq:138}
\phi + {^{112}_{~48}}{\rm Cd} \to {^{112}_{~49}}{\rm In} + e^- + \bar{\nu}_e,
\end{eqnarray}
where ${^{112}_{~48}}{\rm Cd}$ is an atom with a stable nucleus in the
ground state with spin$^{(\rm parity)}$ $J^{\pi} = 0^+$. Since
${^{112}_{~49}}{\rm In}$ is an atom with a nucleus in the ground state
with spin$^{(\rm parity)}$ $J^{\pi} = 1^+$, the reaction
Eq.(\ref{eq:138}) is determined by the Gamow--Teller transition
${^{112}_{~48}}{\rm Cd} \to {^{112}_{~49}}{\rm In}$ for chameleon
energies $E_{\phi} \ge 3.088\,{\rm MeV}$. The calculation of the
threshold energies in weak decay of heavy atoms with the account for
the contribution of the electron shells can be found in
\cite{Ivanov2008}.

We have to note that the atom ${^{112}_{~49}}{\rm In}$ is unstable
under the electron capture (EC) and $\beta^-$ decays with the branches
$56\,\%$ and $44\,\%$, respectively \cite{Tuli2005}. Since the EC
decay of ${^{112}_{~49}}{\rm In}$ is not observable in the experiment
on the chameleon--induced $\beta^-$--decay, one may observe the
$\beta^-$--decay of ${^{112}_{~49}}{\rm In}$. However, the electron
energy spectrum of ${^{112}_{~49}}{\rm In}$ is restricted by the
end--point energy $E_0 = 0.673\,{\rm MeV}$ and can be distinguished
from the energy spectrum of the electron, appearing in the final state
of the reaction Eq.(\ref{eq:138}).

The main background for the chameleon--induced
  $\beta^-$--decay Eq.(\ref{eq:138}) is the reaction
\begin{eqnarray}\label{eq:139}
\nu_e + {^{112}_{~48}}{\rm Cd} \to {^{112}_{~49}}{\rm In} + e^-,
\end{eqnarray}
caused by solar neutrinos with energies $E_{\nu_e} \ge 3.088\,{\rm
  MeV}$. Because of the threshold energy the reaction
Eq.(\ref{eq:139}) can be induced by only the solar ${^8}{\rm B}$ and
{\it hep} neutrinos \cite{PDG2014}. Since the {\it hep}--solar
neutrino flux is approximately 1000 times weaker in comparison to the
${^8}{\rm B}$--solar neutrino flux \cite{PDG2014}, the reaction
Eq.(\ref{eq:139}) should be induced by the ${^8}{\rm B}$--solar
neutrinos.

Concluding our analysis of standard electroweak interactions in the
gravitational theory we would like to discuss the results, obtained
recently by Obukhov {\it et al.}  \cite{Obukhov2014}. There, the
behaviour of the Dirac fermions in the Poincar$\acute{\rm e}$ gauge
gravitational field including a torsion was analysed. The Hamilton
operator of the spin--torsion interaction has been derived
\cite{Obukhov2014}. In a weak gravitational field and torsion field
approximation, which we develop in this paper, such a spin--torsion
interaction takes the form
\begin{eqnarray}\label{eq:140}
{\rm H}_{\rm spin-tors} = - \frac{1}{4}\,(\vec{\Sigma}\cdot \vec{T} +
\gamma^5 T^0),
\end{eqnarray}
where $\vec{\Sigma} = \gamma^0 \vec{\gamma}\gamma^5$ and $\gamma^5 =
i\gamma^0\gamma^1\gamma^2 \gamma^3$ are the Dirac matrices
\cite{Itzykson1980}.  Then, $T^0$ and $\vec{T}$ are the time and
spatial components of the axial torsion vector field $T^{\alpha} =
(T^0, \vec{T}\,)$, defined by
\begin{eqnarray}\label{eq:141}
T^{\alpha} = - \frac{1}{2}\,\varepsilon^{\alpha\beta\mu\nu}\,{\cal T}_{\beta\mu\nu},
\end{eqnarray}
where ${\cal T}_{\beta\mu\nu}$ is the torsion tensor field and
$\varepsilon^{\alpha\beta\mu\nu}$ is the totally antisymmetric
Levi--Civita tensor $\varepsilon^{0123} = 1$
\cite{Itzykson1980}. Using the experimental data \cite{Venema1992} and
\cite{Gemmel2010} on the measurements of the ratio of the nuclear
spin--precession frequencies of the pairs of atoms $({^{199}}{\rm
  Hg},{^{201}}{\rm Hg})$ \cite{Venema1992} and $({^{3}}{\rm
  He},{^{129}}{\rm Xe})$ \cite{Gemmel2010} with nuclear spins and
parities $(J^\pi = \frac{1}{2}^-, J^\pi = \frac{3}{2}^-)$ and $(J^\pi
= \frac{1}{2}^+, J^\pi = \frac{1}{2}^+)$, respectively, Obukhov,
Silenko and Teryaev \cite{Obukhov2014} have found the strong new upper
bound on the absolute value of the torsion axial vector field
$\vec{T}$. They have got
\begin{eqnarray}\label{eq:142}
|\vec{T}\,||\cos\Theta| < 4.7\times 10^{-22}\,{\rm eV}. 
\end{eqnarray}
In the approach, developed in our paper, the tensor torsion field
${\cal T}_{\beta\mu\nu}$ is equal to ${\cal T}_{\beta\mu\nu} =
(\beta/M_{\rm Pl})(g_{\beta\nu}\partial_{\mu}\phi -
g_{\beta\mu}\partial_{\nu}\phi)$ (see Eq.(\ref{eq:7})).  Multiplying
such a tensor torsion field by the totally antisymmetric Levi-Civita
tensor $\varepsilon^{\alpha\beta\mu\nu}$ we get $T^{\alpha} =
0$. Thus, the upper bound of the absolute value of the axial vector
torsion field Eq.(\ref{eq:142}), obtained by Obukhov, Silenko and
Teryaev \cite{Obukhov2014}, does not rule out a possibility for a
torsion field to be induced by the chameleon field as it is proposed
in our paper.

\section{Acknowledgements}

We thank Hartmut Abele for stimulating discussions, Mario Pitschmann
for collaboration at the initial stage of the work and Justin Khoury
for the discussions.  The main results of this paper have been
reported at the International Conference ``Gravitation: 100 years
after GR'' held at Rencontres de Moriond on 21 - 28 March 2015 in La
Thuile, Italy. This work was supported by the Austrian ``Fonds zur
F\"orderung der Wissenschaftlichen Forschung'' (FWF) under the
contract I862-N20.

\end{document}